\documentclass[namedreferences,hyperref,optionalrh]{spr-sola}
\usepackage[utf8]{inputenc}
\usepackage{graphicx,amssymb}        
\usepackage{color}           

\newcommand{\aap}{{\it Astron. Astrophys.}}

\newcommand{\apj}{{\it Astrophys. J.}}
\newcommand{\apjl}{{\it Astrophys. J. Lett.}}

\newcommand{\jgr}{{\it J. Geophys. Res.}}

\newcommand{\solphys}{{\it Sol. Phys.}}
 
\newcommand{\ssr}{{\it Space Sci. Rev.}} 
\chardef\us=`\_
\graphicspath{{./}{figures/}}
\begin{document}

\begin{frontmatter}
\title{The Solar Origin of an Intense Geomagnetic Storm on 2023 December 1st: Successive Slipping and Eruption of Multiple Magnetic Flux Ropes}

\author[addressref={aff1}]{\inits{Z.}\fnm{Zheng}~\snm{Sun}\orcid{0000-0001-5657-7587}}
\author[addressref={aff2,aff3,aff4},email={liting@nao.cas.cn}]{\inits{T.}\fnm{Ting}~\snm{Li}\orcid{0000-0001-6655-1743}}
\author[addressref={aff2,aff3,aff4}]{\inits{Y.}\fnm{Yijun}~\snm{Hou}\orcid{0000-0002-9534-1638}}
\author[addressref={aff1,aff4}]{\inits{H.}\fnm{Hui}~\snm{Tian}\orcid{0000-0002-1369-1758}}
\author[addressref={aff1}]{\inits{Z.}\fnm{Ziqi}~\snm{Wu}\orcid{0000-0002-1349-8720}}
\author[addressref={aff1}]{\inits{K.}\fnm{Ke}~\snm{Li}}
\author[addressref={aff2}]{\inits{Y.}\fnm{Yining}~\snm{Zhang}\orcid{0000-0001-5933-5794}}
\author[addressref={aff5}]{\inits{Z.}\fnm{Zhentong}~\snm{Li}\orcid{0000-0002-4230-2520}}
\author[addressref={aff2}]{\inits{X.}\fnm{Xianyong}~\snm{Bai}\orcid{0000-0003-2686-9153}}
\author[addressref={aff5,aff6}]{\inits{L.}\fnm{Li}~\snm{Feng}\orcid{0000-0003-4655-6939}}
\author[addressref={aff7,aff8}]{\inits{C.}\fnm{Chuan}~\snm{Li}\orcid{0000-0001-7693-4908}}
\author[addressref={aff1}]{\inits{Z.}\fnm{Zhenyong}~\snm{Hou}\orcid{0000-0003-4804-5673}}
\author[addressref={aff9,aff10,aff11}]{\inits{Q.}\fnm{Qiao}~\snm{Song}\orcid{0000-0003-3568-445X}}
\author[addressref={aff9,aff10,aff11}]{\inits{J.}\fnm{Jingsong}~\snm{Wang}}
\author[addressref={aff2,aff3,aff4}]{\inits{G.}\fnm{Guiping}~\snm{Zhou}\orcid{0000-0001-8228-565X}}

\address[id=aff1]{School of Earth and Space Sciences, Peking University, Beijing 100871, People's Republic of China}
\address[id=aff2]{National Astronomical Observatories, Chinese Academy of Sciences, Beijing 100101, People's Republic of China}
\address[id=aff3]{School of Astronomy and Space Science, University of Chinese Academy of Sciences, Beijing 100049, People's Republic of China}
\address[id=aff4]{Key Laboratory of solar activity and space weather, National Space Science Center, Chinese Academy of Sciences, Beijing 100190, People's Republic of China}
\address[id=aff5]{Key Laboratory of Dark Matter and Space Astronomy, Purple Mountain Observatory, Chinese Academy of Sciences, Nanjing 210023, People's Republic of China}
\address[id=aff6]{School of Astronomy and Space Science, University of Science and Technology of China, Hefei 230026, People's Republic of China}
\address[id=aff7]{School of Astronomy and Space Science, Nanjing University, Nanjing 210023, People's Republic of China}
\address[id=aff8]{Key Laboratory for Modern Astronomy and Astrophysics (Nanjing University), Ministry of Education, Nanjing 210023, People's Republic of China}
\address[id=aff9]{National Satellite Meteorological Center (National Centre for Space Weather), China Meteorological Administration, Beijing 100081, People's Republic of China}
\address[id=aff10]{Innovation Center for FengYun Meteorological Satellite (FYSIC), Beijing 100081, People's Republic of China}
\address[id=aff11]{Key Laboratory of Space Weather, National Center for Space Weather, China Meteorological Administration, Beijing 100081, People's Republic of China}

\runningauthor{Sun et al.}
\runningtitle{multiple MFRs}

\begin{abstract}
The solar eruption that occurred on 2023 November 28 (SOL2023-11-28) triggered an intense geomagnetic storm on Earth on 2023 December 1. The associated Earth's auroras manifested at the most southern latitudes in the northern hemisphere observed in the past two decades. In order to explore the profound geoeffectiveness of this event, we conducted a comprehensive analysis of its solar origin to offer potential factors contributing to its impact. Magnetic flux ropes (MFRs) are twisted magnetic structures recognized as significant contributors to coronal mass ejections (CMEs), thereby impacting space weather greatly. In this event, we identified multiple MFRs in the solar active region and observed distinct slipping processes of the three MFRs: MFR1, MFR2, and MFR3. All three MFRs exhibit slipping motions at a speed of 40--137 km s$^{-1}$, extending beyond their original locations. Notably, the slipping of MFR2 extends to $\sim$30 Mm and initiate the eruption of MFR3. Ultimately, MFR1's eruption results in an M3.4-class flare and a CME, while MFR2 and MFR3 collectively produce an M9.8-class flare and another halo CME. This study shows the slipping process in a multi-MFR system, showing how one MFR's slipping can trigger the eruption of another MFR. We propose that the CME--CME interactions caused by multiple MFR eruptions may contribute to the significant geoeffectiveness.
\end{abstract}
\end{frontmatter}

\section{Introduction}
     \label{S-Introduction} 

Magnetic flux ropes (MFRs) are constituted by a collection of helical magnetic field lines winding around a common axis \citep{titov1999basic,krall2001erupting,liu2020magnetic}. Extensive observations and simulations highlight the significant role of MFRs in driving solar flares and coronal mass ejections (CMEs), which have important impacts on the Earth and planetary space environments \citep{inoue2006three,cheng2011observing,song2015evidence,jiang2019reconstruction}. Statistical analysis reveals that approximately half of the CMEs are accompanied by the presence of MFRs \citep{nindos2015common}. Therefore, investigating the detailed mechanisms of the MFR's formation, initiation, and evolution processes is essential for a better understanding of their effects on space weather.

Two prevailing assumptions regarding the formation of an MFR exist. Firstly, MFRs can be pre-existing before an eruption, originated from the convection zone through flux emergence \citep{leka1996evidence,manchester2004eruption,fan2009emergence,torok2014distribution} or from magnetic reconnection preceding the eruption \citep{patsourakos2013direct,chintzoglou2015formation,cheng2023deciphering}. Secondly, MFRs may be formed during the eruption through magnetic reconnection \citep{cheng2011observing,wang2017buildup,liu2018rapid}. An additional scenario involves a pre-existing seed MFR, which continues its buildup as the eruption progresses \citep{cheng2017origin,veronig2018genesis,chen2019observational}. 

Coronal sigmoids, S-shaped magnetic structures above polarity-inversion lines (PILs) observed in soft X-ray and EUV, are often considered MFR progenitors \citep{chen2011coronal,jiang2013magnetohydrodynamic,archontis2014recurrent}. They could either be the MFR itself or sheared arcades providing magnetic conditions for MFR buildup \citep{gibson2006evolving,liu2010sigmoid,savcheva2012photospheric}. Solar filaments, dense and cold plasma suspended in the tenuous and hot corona, can also be regarded as MFRs or sheared arcades \citep{canou2010twisted,li2018two,sun2023observation}. Observations indicate that they can act as seed MFRs and increase poloidal magnetic flux during eruptions \citep{liu2019formation,xie2023magnetic}. There can be simultaneous multiple MFRs in one eruption, which may cause a complicated eruption process \citep{ding2006catastrophic,hou2018eruption,hou2023partial,awasthi2018pre}.

Upon formation, the MFR exhibits distinct magnetic connectivities compared to the surrounding field lines due to their twisted nature. Consequently, the twisted MFR are wrapped by quasi-separatrix layers (QSLs), serving as a boundary between the MFR and the surrounding magnetic field \citep{demoulin1996quasi,demoulin1997quasi,titov1999basic,titov2002theory}. The cross-section between the QSLs and the photosphere, known as the QSL footprints, often manifests a double J shape, with the hook part encircling the MFR footpoints. This characteristic has been validated through numerous simulations and observations \citep{aulanier2012standard,janvier2013standard,zhao2016hooked}. Notably, several studies indicate a correspondence between the flare ribbons and QSL footprints \citep{savcheva2015relation,zhao2016hooked,jiang2018magnetohydrodynamic}.

Three-dimensional (3D) magnetic reconnection occuring in the QSLs is often associated with apparent slipping motion of field lines, known as the ``slipping reconnection" \citep{priest1992magnetic,priest1995three,li2021three}. In the 3D MFR eruption models, overlying sheared arcades can configure into hyperbolic flux tube (HFT) structures beneath the MFR during or before the eruption \citep{aulanier2012standard}. The slipping reconnection within the HFT typically results in continuous brightenings observed at the footprints, often at sub-Alfvénic speeds \citep{aulanier2006slip,janvier2013standard,dudik2014slipping}. This reconnection process enhances the poloidal flux of the MFR, thereby facilitating its buildup. Previous observations and simulations have shown the slipping reconnection in the process of the buildup of an MFR \citep{janvier2013standard,dudik2014slipping,dudik2016slipping,li2014slipping,li2016slipping,li2018three}.

The eruption of MFRs forms the CMEs, which propagate into interplanetary space and are typically referred to as interplanetary CMEs (ICMEs). Some ICMEs directed towards Earth can be geoeffective, leading to geomagnetic storms and aurora on Earth. The disturbance storm time (Dst) and Kp indices are used to quantify the geomagnetic effect. In this study, our focus centers on a solar eruption characterized by two successive solar flares, first an M3.4-class flare followed by an M9.8-class flare, occurring on 2023 November 28. This particular eruption leads to an intense geomagnetic storm  on 2023 December 1st, with a Dst index of $-105$ nT. This event stands out as one of the six intense geomagnetic storms (Dst $\le$ $-100$ nT) recorded since Solar Cycle 25.  Particularly notable is that the auroras associated with this event manifest at the most southern latitudes in the northern hemisphere for the past two decades.
Given the exceptional geoeffectiveness of this event, it is necessary to carry out a thorough analysis of the solar source of the eruption. Our objective is to unravel the possible reasons contributing to its profound geomagnetic effects from the solar observations.
The remainder of this paper is organized as follows. Section 2 describes the observations and data analysis used in our study. In Section 3, we present our observational results in detail. The major findings are thoroughly discussed and summarized in Section 4. Finally, we draw our conclusions in Section 5.

\section{Observations and data analysis} \label{sec:observations}
The Atmospheric Imaging Assembly (AIA) onboard the Solar Dynamics Observatory (SDO) can provide both extreme ultraviolet (EUV) and ultraviolet (UV) images of the Sun since 2010, capturing atmospheric dynamics in a wide range of temperatures from $\log (T) \approx 4.0$ K to $\log (T) \approx 7.0$ K \citep{lemen2012atmospheric}. We analyzed 94~\AA, 131~\AA, 171~\AA, 193~\AA,  and 304~\AA\ data with a spatial resolution of 0.6$^{\prime\prime}$/pixel and a temporal resolution of 12 seconds in our study. UV 1600~\AA\ observations with a resolution of 0.6$^{\prime\prime}$ pixel$^{-1}$ and 24 seconds are also used.
The Advanced Space-based Solar Observatory (ASO-S; \citealt{gan2019advanced}) is the first comprehensive Chinese dedicated solar observatory in space, proposed for the 25th solar maximum. Three payloads are deployed: the Hard X-ray Imager (HXI; \citealt{zhang2019hard}), the Full-disk vector MagnetoGraph (FMG; \citealt{deng2019design}) and the Lyman-$\alpha$ Solar Telescope (LST; \citealt{li2019lyman}). The time cadence varies from 4 seconds in regular observation mode to approximately 0.125 seconds in burst mode. Our study utilizes the full-disk light curve from three total flux monitors (D92, D93, D94). The data is subtracted by the summed count rates of the corresponding three background monitors (D95, D96, D99), focusing on the energy range of 10--50 keV.
The White-light Solar Telescope (WST) onboard LST works in the 360±2 nm waveband and has a field of view of 1.2 solar radii. The images taken by WST have a size of 4608 $\times$ 4608 pixels and the pixel size is $\sim$0.5$^{\prime\prime}$.
FMG can measure the solar photospheric magnetic fields through the Fe \small{I} 532.42 nm line with high spatial and temporal resolutions, providing line-of-sight (LOS) magnetograms.
Solar X-ray Extreme Ultraviolet Imager (X-EUVI; \citealt{chen2022solar}) onboard Fengyun-3E meteorological satellite (FY-3E) can provide EUV images in 195.5 Å, with a cadence of $\sim$14 s and a pixel size of 2$^{\prime\prime}$.5.
The Solar Upper Transition Region Imager (SUTRI; \citealt{bai2023solar}) contributes to our study by providing EUV images at the wavelength of Ne VII 465~\AA, corresponding to a temperature of approximately 0.5 MK \citep{tian2017probing}. SUTRI's capabilities include a spatial resolution of approximately 8$^{\prime\prime}$ and a time resolution of about 30 seconds.
Moreover, H$\alpha$ images from the New Vacuum Solar Telescope (NVST; \citealt{liu2014new}) and the Chinese H$\alpha$ Solar Explorer (CHASE; \citealt{li2022chinese}) are used to show the filaments involved in the event. NVST offers H$\alpha$ images with a field of view (FOV) of 150$^{\prime\prime}$ $\times$ 150$^{\prime\prime}$, a cadence of 45 seconds, and a spatial resolution of 0.165$^{\prime\prime}$ per pixel. Calibration of the data follows the methodology outlined by \citet{cai2022co}. CHASE's H$\alpha$ Imaging Spectrograph (HIS) has a pixel spectral resolution of 0.024~\AA\ and a pixel spatial resolution of 0.52$^{\prime\prime}$.
The Geostationary Operational Environmental Satellite (GOES) can provided soft X-ray (SXR) flux data in 1-8~\AA\ and 0.5-4~\AA, with a cadence of 1 second. 
The white-light observations of CMEs are acquired from the Large Angle Spectrometric Coronagraph (LASCO; \citealt{brueckner1995large}) instrument aboard the Solar and Heliospheric Observatory (SOHO) spacecraft (Brueckner et al., 1995). We utilize LASCO-C3, which offers a FOV spanning from 3 to 32 solar radii, enabling us to observe the CME within the near-Sun corona.
In addition, the Wind satellite \citep{harten1995design} provides magnetic and plasma observations from the L1 point, allowing for detailed monitoring of solar wind properties and magnetic field dynamics in the vicinity of the Earth.

To derive the 3D coronal magnetic field, we utilized the photospheric vector magnetic field data provided by SDO/HMI with a cadence of 720 seconds. A non-linear force-free field (NLFFF) extrapolation was performed using an optimization approach method \citep{wheatland2000optimization, wiegelmann2004optimization, wiegelmann2012should}. To ensure the force-free assumption is satisfied in our input data, we applied the preprocessing method introduced by \citet{wiegelmann2006preprocessing}. The extrapolation box was defined as 740 ${\times}$ 480 ${\times}$ 256, with a grid resolution of 0.5$^{\prime\prime}$. Utilizing the extrapolated field, we computed the twist number $T_w$ employing the procedure developed by \citet{liu2016structure}. The twist number $T_w$ serves as an indicator of the degree of twisting in magnetic field lines, which can help us to identify the MFRs \citep{berger2006writhe}.

\section{Results} \label{sec:results}
On 2023 November 28, NOAA active region (AR) 13500 was positioned at S20$^\circ$W04$^\circ$, slightly south of the center of the solar disk (Figure 1(b)). Figure 1 presents the overview of this AR. The AR exhibited a significant solar eruption, first an M3.4-class flare (SOL2023-11-28T19:07) starting at $\sim$19:07 UT, reaching its peak at $\sim$19:32 UT. Following this event, a subsequent M9.8-class flare (SOL2023-11-28T19:35) occurred, beginning at $\sim$19:35 UT and peaking at $\sim$19:50 UT. Figure 1(a) illustrates the SXR and HXR light curves corresponding to these associated flares.
In panel (c), zoom-in images of the EUV 94~\AA\ wavelength reveal the presence of a sigmoid locating around the PIL, suggesting the existence of a MFR or sheared magnetic arcades. 

Figure 2 presents the NLFFF extrapolation results of the AR 13500 at 18:36 UT, which is approximately half an hour before the eruption. This AR exhibits three distinct MFRs: MFR1 and MFR2 are near each other, with their west ends anchoring the northwest negative-polarity sunspot. The magnetic twist of MFR1 is calculated in the range of 1.0--2.1, and the twist of MFR2 is smaller than MFR1 with its maximum twist of about 1.25. MFR3 is approximately 30 Mm away from MFR2, anchoring in the south negative-polarity sunspot. The twist of the MFR3 is in the range of 1.0--1.63. 
Pre-eruptive observations of the AR are shown in the first row of Figure 3. H$\alpha$ observations in panels (a) and (b) reveal multiple filaments, with particular emphasis on two (outlined by yellow and red dashed lines) located in the northwest. Combining the magnetogram in panel (c), it is found that the locations of two filament are close to those of MFR1 and MFR2 extrapolated in Figure 2. The east end of the extrapolated MFR2 is somewhat north to the observed filament (red dashed lines in Figures 3((a)-(c)), which may imply that the observed filament is composed of both twist magnetic fields and sheared arcades \citep{guo2010coexisting}. These filaments can be considered as the counterparts of the extrapolated MFRs \citep{canou2010twisted,li2018two}. Though there is no associated filament for the extrapolated MFR3, the EUV observations show the presence of high-temperature sheared or twisted loop bundles at the location of MFR3 (Figures 3(d1)-(d3)), which probably correspond to the twisted loop bundles of MFR3 \citep{cheng2011observing,liu2020magnetic}. Panel (c) reveals that the AR possesses a $\beta\gamma\delta$ magnetic configuration, indicative of a complicated magnetic field configuration that tend to erupt as obtained through statistical works \citep{hale1938magnetic,shi1994delta,toriumi2019flare}. Beneath F1 and near the western footpoint of F2, the emergence of numerous discrete magnetic poles of opposite polarity are observed. Motions in the photosphere may lead to magnetic cancellation, making the system unstable and initiating the eruption.

The second to fourth rows of Figure 3 illustrate the initial eruption of MFR1. Brightenings near the location of F1 occur at $\sim$19:13 UT as shown in panels (d1) and (e1). Then the east end of F1 exhibited the apparent slipping motion towards the southeast (yellow arrows in panels (d2)--(f2)). Anchored at the newly-drifted eastern footpoint, a loop bundle (LB1) marked as yellow dots can be observed. This observed feature in 131~\AA\ has been investigated by previous studies and often considered as part of the hot channel, which is one of the indicators for MFRs \citep{cheng2011observing, chintzoglou2015formation, gou2019birth, zheng2021compound}.  At $\sim$19:19 UT, the slipping motion of the east end of LB1 continued with the extension of about 10 Mm (panels (d3)--(f3)). At the same time, another loop bundle LB2 near LB1 occurred, marked as the red dotted line. After that, LB1 and F1 showed an ascending and then erupted (seen in Figure 3(f3) and the online animation), however, LB2 and F2 remained. It can be speculated that the LB1 are associated with F1, which represents an indicator of twisted magnetic field lines of MFR1. On the other hand, LB2 and F2 are probably both associated with MFR2. This slipping motion of MFR during eruptions has been reported by many previous studies \citep{dudik2014slipping,li2014slipping,li2016slipping,wang2017buildup}, indicating the buildup of MFRs.

After the eruption of F1, continual eastward slipping motion started from about 19:21 UT (red arrows in Figure 4). The east end of the slippage extended towards the southeast and passed through the positive-polarity sunspot at the east of the AR (panels (c1)-(c3)). The slippage lasted for about 10 min and showed a sequence of brightening of approximately 30 Mm from the initial eastern end of seed MFR2. Moreover, the AIA observations showed the presence of dense jet-like structures at the base of the east end of the slippage. Combining the observations, we thought that these jet-like structures are generated from the loop bundles of MFR2. In this way, the slipping process is likely to be associated with buildup of MFR2. Finally, the slippage halted near the location of LB3 (marked as green dotted line in panel (b3)), followed by LB3's ascending (Figure 5), indicating that the slippage might initiate the eruption of MFR3.
The slipping motion near LB3 was observed to begin from the eastern footpoint of LB3 (the first row of Figure 5), which depicts a hook-shaped part of the positive-polarity flare ribbon (PR). At the same time, the west negative-polarity flare ribbon (NR) marked by blue arrows in panel (b2) appears. Noteworthy, the slipping of NR was bi-directional, as shown in panels (b1)-(b3). The slippage continued with the flare ribbon elongating southward approximately 70 Mm (panels (a1)-(a4) and (b1)-(b4)). Finally, the LB3 kept ascending and erupted, followed by a flare shown in panel (c4). From the online animation, the bi-directional slipping of NR corresponds to the slipping of PR and hook positive-polarity ribbon (HPR), respectively, which may indicate that this might be associated with both MFR2 and MFR3. Furthermore, the flare region covers the region of both MFR2 and MFR3, which is another evidence of the simultaneous slippage. Ultimately, the simultaneous eruptions of MFR2 and MFR3 contribute to an M9.8-class flare, as depicted in Figure 5(c4).

To characterize the kinetic dynamics of the slipping process and flare ribbon motions, we employed selected slices on the EUV images. Slice A-B in Figure 4(a2) illustrates the slipping process of MFR1 and MFR2, while the corresponding time-distance diagram is presented in Figure 6(a). Initially, MFR1 exhibits slipping motion at a velocity of 106 km s$^{-1}$, shown in the distance from 60 to 30 Mm in panel (a). In this process, the slipping of MFR2 is covered by that of MFR1, making it difficult to distinguish. Following MFR1's slipping and subsequent eruption, MFR2 continues slipping southeastward at 40 km s$^{-1}$ from 30 to 10 Mm in panel (a). This slipping extends toward the vicinity of MFR3, potentially initiating the slippage of MFR3.
Next, slice C-D in Figure 5(a4) is placed to observe the southward slipping of MFR3, and the associated time-distance diagram is depicted in Figure 6(b). The starting time of MFR3's slipping process is corresponding to when the sequential brightening of MFR2 reaches near MFR3, suggesting that MFR3's slippage may be triggered by MFR2. Then, the east end of MFR3 moves southward at the speed of $\sim$137 km s$^{-1}$ to a distance of $\sim$70 Mm.
Slice E-F in Figure 5(b4) elucidates the slipping direction of the negative-polarity ribbon of the double flare ribbon, and the corresponding time-distance diagram is shown in Figure 6(c). Bi-directional brightenings indicate simultaneous eruptions of MFR2 and MFR3. The brightening of MFR2 occurs earlier than that of MFR3, indicating that the buildup process of MFR2 is earlier. The slipping speeds of the PR are measured at 26 km s$^{-1}$ towards the south, which probably corresponds to the west end of MFR2, and 51 km s$^{-1}$ towards the north corresponding to the west end of the MFR3.
Further examination of the time-distance diagram of Slice G-H in Figure 5(b4), represented in Figure 6(d), shows the separation of the double flare ribbons. Initially, around 19:15 UT, PR begins to separate due to the M3.4-class flare at the speed of 16 km s$^{-1}$. Subsequently, around 19:35 UT, the negative-polarity ribbon starts to separate following the M9.8-class flare at a velocity of 18 km s$^{-1}$. Notably, during this period, the northern ribbon remains relatively stable, exhibiting minimal movement perpendicular to the PIL.

Fine structures at the newly-formed eastern footpoint of MFR3 are highlighted within the white square in Figure 5(b4). The temporal evolution of this zoomed-in area is detailed in Figure 7. Analysis of AIA 304~\AA\ and SUTRI 465~\AA\ observations reveals that the elongation of the HPR halts at $\sim$19:42 UT. Subsequently, the end of the flare ribbon undergoes a transformation to a curved shape, as observed in panels (a3) and (d3). Approximately 7 minutes later, it evolves into a pentagram shape (panels (a4),(a5),(b4), and (b5)). This observed evolution process may be attributed to the reconnection between the MFR and surrounding magnetic field lines, leading to the drifting of footpoints as predicted in \citet{aulanier2019drifting}.

Large-scale EUV waves generated by the eruptions are shown in Figure 8. Panel (a) presents the EUV waves accompanying the M3.4-class flare, which is triggered by the eruption of MFR1. The wave propagates northward at a speed of 980 km s$^{-1}$ with an angular width of approximately 120$^\circ$. Panels (b) and (c) depict the waves generated by the M9.8-class flare, which results from the eruptions of MFR2 and MFR3. Notably, we observe the expansion of loops from MFR3, indicated by pink cross signs. The expansion of loops serves as an indicator for the ascent of MFR3, with observations revealing a slow rise at 45 km s$^{-1}$ followed by a rapid increase to 390 km s$^{-1}$. This slow-to-fast transition aligns with previous observations of MFR dynamics \citep{zhang2012observation,kliem2014slow,cheng2023deciphering}. In front of the loop expansion, an EUV wave propagates southeastward at a speed of 1080 km s$^{-1}$. Additionally, another wavefront propagates northward at a speed of 1050 km s$^{-1}$ (panel (c)), which is associated with the eruption of MFR2. Such a speed of these EUV waves should be explained as fast magnetoacoustic waves or shock waves \citep{patsourakos2012nature,warmuth2015large,sun2022cross}.

Figure 9 displays white-light LASCO observations of the CMEs. LASCO captured two CMEs (CME1 and CME2) during this event, both documented by the SEEDS catalog\footnote{\url{http://spaceweather.gmu.edu/seeds/}}.
At $\sim$21:30 UT, CME1 is observed propagating westward at a speed projected in the plane of sky of 535 km s$^{-1}$. This CME is associated with the M3.4-class flare triggered by the eruption of MFR1. Approximately 24 minutes later, CME2 originates initially at the south limb of the Sun (panel (b)), evolving subsequently into a halo-CME (panels (c) and (d)), with a projected speed of 2227 km s$^{-1}$ as recorded by SEEDS.

The Wind observations are depicted in Figure 10, where data is averaged over one-minute intervals in the Geocentric Solar Ecliptic (GSE) coordinate system. Analysis of the radial velocity data (panel (c)) reveals a background solar wind speed of approximately 350 km s$^{-1}$. A notable sudden increase at $\sim$23:29 UT on November 30th (yellow dashed lines) indicates the arrival of a fast shock, followed by an ICME. After carefully checking of the EUV and coronagraph data, we found that there are two significant halo CMEs on November 27th and 28th, with the first one being a quiescent filament eruption on November 27th and the second one being the eruption in this study. The in-situ data on November 30th and December 1st only show two significant shocks, which are presented in Figure 10. So we could infer that these two significant shocks correspond to the two halo CMEs, respectively. As a result, the first shock should originate from the halo CME caused by a quiescent filament eruption on November 27th, the day prior to the analyzed eruption.
After approximately 9 hours, another fast shock (orange dashed lines) arrived at $\sim$08:48 UT on December 1st. This ICME corresponds to the eruption under analysis. The main body of this ICME likely corresponds to the halo CME2 in Figure 9, however, we cannot exclude the possibility that CME2 caught up with CME1 and merged \citep{li2012twin,lugaz2017interaction}.
Following the ICME shock-sheath region, the magnetic cloud (MC) passes, identified by systematic variations in the magnetic field components, indicative of a strong magnetic field associated with an MFR-like structure. The Bz component changes from negative to positive sign. Based on these observations, we identified the time intervals for the MC as spanning from $\sim$19:48 UT on December 1st to $\sim$01:24 UT on December 2nd. Besides the clear MFR structure, there is a complex structure after December 3rd 00:00 UT, also with a magnetic component reversal. This might be a possible signal of another MFR structure.

Figure 11 presents hourly-averaged in-situ observations of geomagnetic conditions. Upon encountering the first ICME shock (yellow dashed lines), the Bz component increases to approximately $-8$ nT (where minus indicates southward). Subsequently, upon the arrival of the second shock (orange dashed lines) produced by the eruption under analysis, the Bz component further increases to around $-23$ nT. Such a significant southward magnetic field induces a rapid decrease in the Dst index, with the minimum reaching $-105$ nT, classifying the event as an intense geomagnetic storm (Dst $\le$ $-100$ nT; \citealt{gonzalez1994geomagnetic,echer2008interplanetary,lakhina2016geomagnetic}).
Following the main phase of the geomagnetic storm, there is a long-duration recovery phase. Additionally, this event leads to an increase in the Kp index (panel (c)) to 6.7, categorized as G2 according to NOAA Space Weather Scales. Furthermore, this event triggers low-latitude auroras in the northern hemisphere, which falls beyond the scope of this paper.

\section{Discussion} \label{sec:discussion}
This paper investigates a solar eruption that resulted in an intense geomagnetic storm and a low-latitude aurora. The eruption involved three erupting MFRs, two flares and two CMEs. Throughout this event, we observed three slipping process during the buildup of the three MFRs. The possible scenario is as follows: MFR1 initiates its ascent, accompanied by eastward slipping of the its east end at the speed of 106 km s$^{-1}$. Subsequently, MFR1 erupts, leading to the formation of the M3.4-class flare and CME1. Meanwhile, MFR2 undergoes slipping towards the southeast at 40 km s$^{-1}$. As MFR2 slips approximately 30 Mm, it approaches the vicinity of MFR3, triggering the initiation of MFR3. MFR3 then begins slipping, with its flare ribbon elongating southward at the speed of 137 km s$^{-1}$. Ultimately, the eruptions of MFR2 and MFR3 contribute both to the M9.8-class flare and CME2.

Slipping motions in the MFR eruptions have been simulated and observed in many works \citep{janvier2013standard,dudik2014slipping,dudik2016slipping,li2014slipping,li2016slipping}. In the 3D extensions to the standard CSHKP model, MFRs are typically anchored at the hooks of double-J QSLs. There are numerous sheared arcades overlying the MFR, with one end anchored at the hook part of one J-shaped QSL and the other end at the straight part of the other J-shaped QSL (see in Fig.5 in \citealt{aulanier2012standard}). During the pre-eruption or eruption phases of MFRs, the sheared arcades can form a HFT structure beneath the MFR and reconnect with each other. This reconnection process converts two sheared field lines into a magnetic flux rope line and a post-flare loop, thereby increasing the poloidal magnetic flux of the MFR and contributing to its buildup. The continuous slip-running reconnection of these field lines leads to observable slipping motions in the flare ribbons during the buildup of the MFR \citep{janvier2013standard}. The buildup process corresponds to the scenario where a pre-existing seed MFR accumulates poloidal magnetic flux during the eruption phase. Our results reveal extrapolated MFR structures prior to the eruption and the subsequent buildup process, in agreement with the seed MFR eruption scenario.

Numerous observational case studies have confirmed the presence of slipping motions before and during MFR eruptions \citep{dudik2014slipping,dudik2016slipping,li2014slipping,li2016slipping,li2018three}. These studies have identified continuous brightening along the flare ribbons or QSLs during the buildup process of MFR eruptions. Despite many case studies demonstrating the slipping motions of individual MFR eruptions, few studies have investigated the slipping motions in multi-MFR systems. 
In our event, we observed the slipping motions of each MFR, providing observational supplements for the three-dimensional reconnection in the MFR eruption. Additionally, the multi-MFR eruption is commonly explained as the reduction of the restraining forces above \citep{schmieder2013solar,dhakal2018study,green2018origin}. However, in our event, the end of MFR2 was elongated to the vicinity of another flux rope MFR3 through the slipping process, which plausibly triggered the eruption of MFR3. Our observations imply that the slipping reconnection plays a key role in the multi-MFR eruptions.

During the slipping process, we observed that brightenings not only occurred in the immediate vicinity of the MFRs, but extends in length to a distance of up to 70 Mm. This phenomenon can also be clearly identified in \citet{dudik2014slipping}, where an elongation of the flare ribbon is observed. The sequential brightening should occur along the hook part of the QSLs (i.e. the footpoints of the MFR), and thus, the direction and distance of the slipping should be associated with the specific configuretion of MFRs \citep{aulanier2013standard,li2014slipping}. 
\citet{priest2017flux} proposed a scenario for the buildup of MFR suggesting a two-step process: first, the 3D zipper reconnection of sheared arcades forms an MFR along the PIL, followed by quasi-2D main phase reconnection of unsheared flux. The former corresponds to the parallel extension of conjugate flare ribbons in the same direction, while the latter corresponds to the separation of double flare ribbons. However, in our event, we observed that the extension of conjugate flare ribbons HPR and NR occurs in opposite directions (see Figure 5), suggesting that slipping reconnection rather than zipper reconnection is involved in this event.

Numerous observations indicate that some intense geomagnetic storms are linked with multiple CMEs \citep{farrugia2006survey,zhang2007solar,liu2014observations}. CMEs in the near-Sun corona or interplanetary space can interact with each other, termed the CME--CME interaction. Studies have highlighted the significant geoeffectiveness of CME--CME interactions \citep{lugaz2017interaction,joshi2018major,scolini2020cme}. The long-duration driving and compressed magnetic fields resulting from CME--CME interactions can contribute to intense geomagnetic storms \citep{lugaz2017interaction}.
In our event, the halo CME2 includes eruptions possibly from both MFR2 and MFR3, indicating that CME2 itself is a product of CME--CME interaction. Moreover, CME1 preceding CME2 also suggests the possibility of another CME--CME interaction. Consequently, CME--CME interactions among these three MFRs' eruptions may significantly contribute to the observed geomagnetic effects. Additionally, when the ICME associated with this eruption reaches Earth, it encounters the ICME produced on the day before (Figure 10), potentially compressing the prior ICME and further enhancing geoeffectiveness.
Ongoing investigations with a focus on understanding how these CME--CME interactions influence magnetosphere is crucial for the observed profound geoeffectiveness, which is beyond the scope of this paper.

\section{Conclusion} \label{sec:conclusion}
This study investigated a multi-MFR eruption that occurred on 2023 November 28th, involving MFR1, MFR2, and MFR3. MFR1 led to the formation of an M3.4-class flare and CME1, while the eruptions of MFR2 and MFR3 collectively contributed to an M9.8-class flare and CME2. We observed distinct slipping processes during the buildup of the three MFRs. All three MFRs exhibited slipping motions, extending beyond their original locations. Notably, the slipping of MFR2 extends $\sim$30 Mm and plausibly initiate the eruption of MFR3. To our knowledge, this is the first study to display successive slipping process in a multi-MFR system, showing how one MFR's slipping can trigger the slippage and eruption of another MFR.

The eruption resulted in an intense geomagnetic storm, with a Dst index reaching $-105$ nT on December 1st, 2023. This event stands out as one of the six intense geomagnetic storms recorded since Solar Cycle 25. Notably, the associated aurora reached the most southern latitudes in the northern hemisphere observed in the past two decades.
Through analysis of the solar origin of the eruption, we consider that CME--CME interactions, triggered by multiple MFR eruptions, may be significant contributors to the observed geomagnetic effects. Further studies focusing on ICMEs and Earth's magnetosphere are necessary to elucidate the detailed mechanisms behind the intense geomagnetic storm and the low-altitude aurora associated with this event.

\begin{acks}
We thank the referee and the editor for their careful work and thoughtful suggestions. We also thank Prof. Shuo Yao and Dr. Wensi Wang for fruitful discussions. The ASO-S mission is supported by the Strategic Priority Research Program on Space Science, Chinese Academy of Sciences. SUTRI is a collaborative project conducted by the National Astronomical Observatories of CAS, Peking University, Tongji University, Xi'an Institute of Optics and Precision Mechanics of CAS and the Innovation Academy for Microsatellites of CAS. The CHASE mission is supported by China National Space Administration (CNSA). AIA is a payload onboard \emph{SDO}, a mission of NASA's Living With a Star Program. We thank the SDO, Wind, SOHO/LASCO, FY-3E/X-EUVI, SUTRI, CHASE, ASO-S, and NVST for providing data. This work is supported by the Strategic Priority Research Program of the Chinese Academy of Sciences (Grant No. XDB0560000 and XDB41000000), the National Key R\&D Program of China (2019YFA0405000,2021YFA1600500,2022YFF0503800), the National Natural Science Foundations of China (11973056, 12222306, and 12273060, 12333009), and the Youth Innovation Promotion Association CAS (2023063).
\end{acks}


\bibliographystyle{spr-mp-sola}
\bibliography{ref}  

\begin{thebibliography}{98}
\ifx\bisbn     \undefined \def\bisbn  #1{ISBN #1}\fi
\ifx\binits    \undefined \def\binits#1{#1}\fi
\ifx\bauthor   \undefined \def\bauthor#1{#1}\fi
\ifx\batitle   \undefined \def\batitle#1{#1}\fi
\ifx\bjtitle   \undefined \def\bjtitle#1{\textit{#1}}\fi
\ifx\bvolume   \undefined \def\bvolume#1{\textbf{#1}}\fi
\ifx\byear     \undefined \def\byear#1{#1}\fi
\ifx\bissue    \undefined \def\bissue#1{#1}\fi
\ifx\bfpage    \undefined \def\bfpage#1{#1}\fi
\ifx\blpage    \undefined \def\blpage #1{#1}\fi
\ifx\burl      \undefined \def\burl#1{#1}\fi
\ifx\href      \undefined \def\href#1#2{#2}\fi
\ifx\betal     \undefined \def\betal{et al.}\fi
\ifx\bctitle   \undefined \def\bctitle#1{#1}\fi
\ifx\beditor   \undefined \def\beditor#1{#1}\fi
\ifx\bbtitle   \undefined \def\bbtitle#1{\textit{#1}}\fi
\ifx\bedition  \undefined \def\bedition#1{#1}\fi
\ifx\bseriesno \undefined \def\bseriesno#1{\textbf{#1}}\fi
\ifx\blocation \undefined \def\blocation#1{#1}\fi
\ifx\bsertitle \undefined \def\bsertitle#1{\textit{#1}}\fi
\ifx\bsnm      \undefined \def\bsnm#1{#1}\fi
\ifx\bsuffix   \undefined \def\bsuffix#1{#1}\fi
\ifx\bparticle \undefined \def\bparticle#1{#1}\fi
\ifx\barticle  \undefined \def\barticle#1{}\fi
\ifx\binstitute  \undefined \def\binstitute#1{#1}\fi
\ifx\bpublisher  \undefined \def\bpublisher#1{#1}\fi
\ifx\doiurl    \undefined \def\doiurl#1{\href{#1}{DOI}}\fi
\makeatletter
\def\safeHref#1#2#3{\in@{http}{#2}\ifin@\href{#2}{#3}\else\href{#1#2}{#3}\fi}
\makeatother
\ifx\adsurl    \undefined
  \def\adsurl#1{\safeHref{https://ui.adsabs.harvard.edu/abs/}{#1}{ADS}}\fi
\ifx\arxivurl  \undefined
  \def\arxivurl#1{\safeHref{http://arxiv.org/abs/}{#1}{arXiv}}\fi
\ifx\botherref \undefined \def\botherref#1{}\fi
\ifx\url       \undefined \def\url#1{#1}\fi
\ifx\bchapter  \undefined \def\bchapter#1{}\fi
\ifx\bbook     \undefined \def\bbook#1{}\fi
\ifx\bcomment  \undefined \def\bcomment#1{#1}\fi
\ifx\oauthor   \undefined \def\oauthor#1{#1}\fi
\ifx\citeauthoryear \undefined\def \citeauthoryear#1{#1}\fi
\def\endbibitem {}
\ifx\bconflocation  \undefined \def\bconflocation#1{#1} \fi

\bibitem[\protect\citeauthoryear{{Archontis}, {Hood}, and
  {Tsinganos}}{2014}]{archontis2014recurrent}
\begin{barticle}
\bauthor{\bsnm{{Archontis}}, \binits{V.}},
\bauthor{\bsnm{{Hood}}, \binits{A.W.}},
\bauthor{\bsnm{{Tsinganos}}, \binits{K.}}:
\byear{2014},
\batitle{{Recurrent Explosive Eruptions and the ``Sigmoid-to-arcade''
  Transformation in the Sun Driven by Dynamical Magnetic Flux Emergence}}.
\bjtitle{\apjl}
\bvolume{786},
\bfpage{L21}.
\doiurl{https://doi.org/10.1088/2041-8205/786/2/L21}.
\adsurl{2014ApJ...786L..21A}.
\end{barticle}
\endbibitem

\bibitem[\protect\citeauthoryear{{Aulanier} and
  {Dud{\'\i}k}}{2019}]{aulanier2019drifting}
\begin{barticle}
\bauthor{\bsnm{{Aulanier}}, \binits{G.}},
\bauthor{\bsnm{{Dud{\'\i}k}}, \binits{J.}}:
\byear{2019},
\batitle{{Drifting of the line-tied footpoints of CME flux-ropes}}.
\bjtitle{\aap}
\bvolume{621},
\bfpage{A72}.
\doiurl{https://doi.org/10.1051/0004-6361/201834221}.
\adsurl{2019A&A...621A..72A}.
\end{barticle}
\endbibitem

\bibitem[\protect\citeauthoryear{{Aulanier}, {Janvier}, and
  {Schmieder}}{2012}]{aulanier2012standard}
\begin{barticle}
\bauthor{\bsnm{{Aulanier}}, \binits{G.}},
\bauthor{\bsnm{{Janvier}}, \binits{M.}},
\bauthor{\bsnm{{Schmieder}}, \binits{B.}}:
\byear{2012},
\batitle{{The standard flare model in three dimensions. I. Strong-to-weak shear
  transition in post-flare loops}}.
\bjtitle{\aap}
\bvolume{543},
\bfpage{A110}.
\doiurl{https://doi.org/10.1051/0004-6361/201219311}.
\adsurl{2012A&A...543A.110A}.
\end{barticle}
\endbibitem

\bibitem[\protect\citeauthoryear{Aulanier et~al.}{2006}]{aulanier2006slip}
\begin{barticle}
\bauthor{\bsnm{Aulanier}, \binits{G.}},
\bauthor{\bsnm{Pariat}, \binits{E.}},
\bauthor{\bsnm{Demoulin}, \binits{P.}},
\bauthor{\bsnm{DeVore}, \binits{C.}}:
\byear{2006},
\batitle{Slip-running reconnection in quasi-separatrix layers}.
\bjtitle{Solar Physics}
\bvolume{238},
\bfpage{347}.
\end{barticle}
\endbibitem

\bibitem[\protect\citeauthoryear{{Aulanier}
  et~al.}{2013}]{aulanier2013standard}
\begin{barticle}
\bauthor{\bsnm{{Aulanier}}, \binits{G.}},
\bauthor{\bsnm{{D{\'e}moulin}}, \binits{P.}},
\bauthor{\bsnm{{Schrijver}}, \binits{C.J.}},
\bauthor{\bsnm{{Janvier}}, \binits{M.}},
\bauthor{\bsnm{{Pariat}}, \binits{E.}},
\bauthor{\bsnm{{Schmieder}}, \binits{B.}}:
\byear{2013},
\batitle{{The standard flare model in three dimensions. II. Upper limit on
  solar flare energy}}.
\bjtitle{\aap}
\bvolume{549},
\bfpage{A66}.
\doiurl{https://doi.org/10.1051/0004-6361/201220406}.
\adsurl{2013A&A...549A..66A}.
\end{barticle}
\endbibitem

\bibitem[\protect\citeauthoryear{{Awasthi} et~al.}{2018}]{awasthi2018pre}
\begin{barticle}
\bauthor{\bsnm{{Awasthi}}, \binits{A.K.}},
\bauthor{\bsnm{{Liu}}, \binits{R.}},
\bauthor{\bsnm{{Wang}}, \binits{H.}},
\bauthor{\bsnm{{Wang}}, \binits{Y.}},
\bauthor{\bsnm{{Shen}}, \binits{C.}}:
\byear{2018},
\batitle{{Pre-eruptive Magnetic Reconnection within a Multi-flux-rope System in
  the Solar Corona}}.
\bjtitle{\apj}
\bvolume{857},
\bfpage{124}.
\doiurl{https://doi.org/10.3847/1538-4357/aab7fb}.
\adsurl{2018ApJ...857..124A}.
\end{barticle}
\endbibitem

\bibitem[\protect\citeauthoryear{{Bai} et~al.}{2023}]{bai2023solar}
\begin{barticle}
\bauthor{\bsnm{{Bai}}, \binits{X.}},
\bauthor{\bsnm{{Tian}}, \binits{H.}},
\bauthor{\bsnm{{Deng}}, \binits{Y.}},
\bauthor{\bsnm{{Wang}}, \binits{Z.}},
\bauthor{\bsnm{{Yang}}, \binits{J.}},
\bauthor{\bsnm{{Zhang}}, \binits{X.}},
\bauthor{\bsnm{{Zhang}}, \binits{Y.}},
\bauthor{\bsnm{{Qi}}, \binits{R.}},
\bauthor{\bsnm{{Wang}}, \binits{N.}},
\bauthor{\bsnm{{Gao}}, \binits{Y.}},
\bauthor{\bsnm{{Yu}}, \binits{J.}},
\bauthor{\bsnm{{He}}, \binits{C.}},
\bauthor{\bsnm{{Shen}}, \binits{Z.}},
\bauthor{\bsnm{{Shen}}, \binits{L.}},
\bauthor{\bsnm{{Guo}}, \binits{S.}},
\bauthor{\bsnm{{Hou}}, \binits{Z.}},
\bauthor{\bsnm{{Ji}}, \binits{K.}},
\bauthor{\bsnm{{Bi}}, \binits{X.}},
\bauthor{\bsnm{{Duan}}, \binits{W.}},
\bauthor{\bsnm{{Yang}}, \binits{X.}},
\bauthor{\bsnm{{Lin}}, \binits{J.}},
\bauthor{\bsnm{{Hu}}, \binits{Z.}},
\bauthor{\bsnm{{Song}}, \binits{Q.}},
\bauthor{\bsnm{{Yang}}, \binits{Z.}},
\bauthor{\bsnm{{Chen}}, \binits{Y.}},
\bauthor{\bsnm{{Qiao}}, \binits{W.}},
\bauthor{\bsnm{{Ge}}, \binits{W.}},
\bauthor{\bsnm{{Li}}, \binits{F.}},
\bauthor{\bsnm{{Jin}}, \binits{L.}},
\bauthor{\bsnm{{He}}, \binits{J.}},
\bauthor{\bsnm{{Chen}}, \binits{X.}},
\bauthor{\bsnm{{Zhu}}, \binits{X.}},
\bauthor{\bsnm{{He}}, \binits{J.}},
\bauthor{\bsnm{{Shi}}, \binits{Q.}},
\bauthor{\bsnm{{Liu}}, \binits{L.}},
\bauthor{\bsnm{{Li}}, \binits{J.}},
\bauthor{\bsnm{{Xu}}, \binits{D.}},
\bauthor{\bsnm{{Liu}}, \binits{R.}},
\bauthor{\bsnm{{Li}}, \binits{T.}},
\bauthor{\bsnm{{Feng}}, \binits{Z.}},
\bauthor{\bsnm{{Wang}}, \binits{Y.}},
\bauthor{\bsnm{{Fan}}, \binits{C.}},
\bauthor{\bsnm{{Liu}}, \binits{S.}},
\bauthor{\bsnm{{Guo}}, \binits{S.}},
\bauthor{\bsnm{{Sun}}, \binits{Z.}},
\bauthor{\bsnm{{Wu}}, \binits{Y.}},
\bauthor{\bsnm{{Li}}, \binits{H.}},
\bauthor{\bsnm{{Yang}}, \binits{Q.}},
\bauthor{\bsnm{{Ye}}, \binits{Y.}},
\bauthor{\bsnm{{Gu}}, \binits{W.}},
\bauthor{\bsnm{{Wu}}, \binits{J.}},
\bauthor{\bsnm{{Zhang}}, \binits{Z.}},
\bauthor{\bsnm{{Yu}}, \binits{Y.}},
\bauthor{\bsnm{{Ye}}, \binits{Z.}},
\bauthor{\bsnm{{Sheng}}, \binits{P.}},
\bauthor{\bsnm{{Wang}}, \binits{Y.}},
\bauthor{\bsnm{{Li}}, \binits{W.}},
\bauthor{\bsnm{{Huang}}, \binits{Q.}},
\bauthor{\bsnm{{Zhang}}, \binits{Z.}}:
\byear{2023},
\batitle{{The Solar Upper Transition Region Imager (SUTRI) Onboard the
  SATech-01 Satellite}}.
\bjtitle{Research in Astronomy and Astrophysics}
\bvolume{23},
\bfpage{065014}.
\doiurl{https://doi.org/10.1088/1674-4527/accc74}.
\adsurl{2023RAA....23f5014B}.
\end{barticle}
\endbibitem

\bibitem[\protect\citeauthoryear{Berger and Prior}{2006}]{berger2006writhe}
\begin{barticle}
\bauthor{\bsnm{Berger}, \binits{M.A.}},
\bauthor{\bsnm{Prior}, \binits{C.}}:
\byear{2006},
\batitle{The writhe of open and closed curves}.
\bjtitle{Journal of Physics A: Mathematical and General}
\bvolume{39},
\bfpage{8321}.
\doiurl{https://doi.org/10.1088/0305-4470/39/26/005}.
\end{barticle}
\endbibitem

\bibitem[\protect\citeauthoryear{{Brueckner} et~al.}{1995}]{brueckner1995large}
\begin{barticle}
\bauthor{\bsnm{{Brueckner}}, \binits{G.E.}},
\bauthor{\bsnm{{Howard}}, \binits{R.A.}},
\bauthor{\bsnm{{Koomen}}, \binits{M.J.}},
\bauthor{\bsnm{{Korendyke}}, \binits{C.M.}},
\bauthor{\bsnm{{Michels}}, \binits{D.J.}},
\bauthor{\bsnm{{Moses}}, \binits{J.D.}},
\bauthor{\bsnm{{Socker}}, \binits{D.G.}},
\bauthor{\bsnm{{Dere}}, \binits{K.P.}},
\bauthor{\bsnm{{Lamy}}, \binits{P.L.}},
\bauthor{\bsnm{{Llebaria}}, \binits{A.}},
\bauthor{\bsnm{{Bout}}, \binits{M.V.}},
\bauthor{\bsnm{{Schwenn}}, \binits{R.}},
\bauthor{\bsnm{{Simnett}}, \binits{G.M.}},
\bauthor{\bsnm{{Bedford}}, \binits{D.K.}},
\bauthor{\bsnm{{Eyles}}, \binits{C.J.}}:
\byear{1995},
\batitle{{The Large Angle Spectroscopic Coronagraph (LASCO)}}.
\bjtitle{\solphys}
\bvolume{162},
\bfpage{357}.
\doiurl{https://doi.org/10.1007/BF00733434}.
\adsurl{1995SoPh..162..357B}.
\end{barticle}
\endbibitem

\bibitem[\protect\citeauthoryear{{Cai} et~al.}{2022}]{cai2022co}
\begin{barticle}
\bauthor{\bsnm{{Cai}}, \binits{Y.-F.}},
\bauthor{\bsnm{{Yang}}, \binits{X.}},
\bauthor{\bsnm{{Xiang}}, \binits{Y.-Y.}},
\bauthor{\bsnm{{Yan}}, \binits{X.-L.}},
\bauthor{\bsnm{{Jin}}, \binits{Z.-Y.}},
\bauthor{\bsnm{{Liu}}, \binits{H.}},
\bauthor{\bsnm{{Ji}}, \binits{K.-F.}}:
\byear{2022},
\batitle{{The Co-alignment of Winged H{\ensuremath{\alpha}} Data Observed by
  the New Vacuum Solar Telescope}}.
\bjtitle{Research in Astronomy and Astrophysics}
\bvolume{22},
\bfpage{065010}.
\doiurl{https://doi.org/10.1088/1674-4527/ac69b9}.
\adsurl{2022RAA....22f5010C}.
\end{barticle}
\endbibitem

\bibitem[\protect\citeauthoryear{{Canou} and {Amari}}{2010}]{canou2010twisted}
\begin{barticle}
\bauthor{\bsnm{{Canou}}, \binits{A.}},
\bauthor{\bsnm{{Amari}}, \binits{T.}}:
\byear{2010},
\batitle{{A Twisted Flux Rope as the Magnetic Structure of a Filament in Active
  Region 10953 Observed by Hinode}}.
\bjtitle{\apj}
\bvolume{715},
\bfpage{1566}.
\doiurl{https://doi.org/10.1088/0004-637X/715/2/1566}.
\adsurl{2010ApJ...715.1566C}.
\end{barticle}
\endbibitem

\bibitem[\protect\citeauthoryear{Chen et~al.}{2022}]{chen2022solar}
\begin{barticle}
\bauthor{\bsnm{Chen}, \binits{B.}},
\bauthor{\bsnm{Zhang}, \binits{X.-X.}},
\bauthor{\bsnm{He}, \binits{L.-P.}},
\bauthor{\bsnm{Song}, \binits{K.-F.}},
\bauthor{\bsnm{Liu}, \binits{S.-J.}},
\bauthor{\bsnm{Ding}, \binits{G.-X.}},
\bauthor{\bsnm{Dun}, \binits{J.-P.}},
\bauthor{\bsnm{Li}, \binits{J.-W.}},
\bauthor{\bsnm{Li}, \binits{Z.-H.}},
\bauthor{\bsnm{Guo}, \binits{Q.-F.}}, \betal:
\byear{2022},
\batitle{Solar X-ray and EUV imager on board the FY-3E satellite}.
\bjtitle{Light: Science \& Applications}
\bvolume{11},
\bfpage{329}.
\doiurl{https://doi.org/10.1038/s41377-022-01023-z}.
\end{barticle}
\endbibitem

\bibitem[\protect\citeauthoryear{{Chen} et~al.}{2019}]{chen2019observational}
\begin{barticle}
\bauthor{\bsnm{{Chen}}, \binits{H.}},
\bauthor{\bsnm{{Yang}}, \binits{J.}},
\bauthor{\bsnm{{Ji}}, \binits{K.}},
\bauthor{\bsnm{{Duan}}, \binits{Y.}}:
\byear{2019},
\batitle{{Observational Analysis on the Early Evolution of a CME Flux Rope:
  Preflare Reconnection and Flux Rope{\textquoteright}s Footpoint Drift}}.
\bjtitle{\apj}
\bvolume{887},
\bfpage{118}.
\doiurl{https://doi.org/10.3847/1538-4357/ab527e}.
\adsurl{2019ApJ...887..118C}.
\end{barticle}
\endbibitem

\bibitem[\protect\citeauthoryear{{Chen}}{2011}]{chen2011coronal}
\begin{barticle}
\bauthor{\bsnm{{Chen}}, \binits{P.F.}}:
\byear{2011},
\batitle{{Coronal Mass Ejections: Models and Their Observational Basis}}.
\bjtitle{Living Reviews in Solar Physics}
\bvolume{8},
\bfpage{1}.
\doiurl{https://doi.org/10.12942/lrsp-2011-1}.
\adsurl{2011LRSP....8....1C}.
\end{barticle}
\endbibitem

\bibitem[\protect\citeauthoryear{{Cheng}, {Guo}, and
  {Ding}}{2017}]{cheng2017origin}
\begin{barticle}
\bauthor{\bsnm{{Cheng}}, \binits{X.}},
\bauthor{\bsnm{{Guo}}, \binits{Y.}},
\bauthor{\bsnm{{Ding}}, \binits{M.}}:
\byear{2017},
\batitle{{Origin and Structures of Solar Eruptions I: Magnetic Flux Rope}}.
\bjtitle{Science China Earth Sciences}
\bvolume{60},
\bfpage{1383}.
\doiurl{https://doi.org/10.1007/s11430-017-9074-6}.
\adsurl{2017ScChD..60.1383C}.
\end{barticle}
\endbibitem

\bibitem[\protect\citeauthoryear{{Cheng} et~al.}{2011}]{cheng2011observing}
\begin{barticle}
\bauthor{\bsnm{{Cheng}}, \binits{X.}},
\bauthor{\bsnm{{Zhang}}, \binits{J.}},
\bauthor{\bsnm{{Liu}}, \binits{Y.}},
\bauthor{\bsnm{{Ding}}, \binits{M.D.}}:
\byear{2011},
\batitle{{Observing Flux Rope Formation During the Impulsive Phase of a Solar
  Eruption}}.
\bjtitle{\apjl}
\bvolume{732},
\bfpage{L25}.
\doiurl{https://doi.org/10.1088/2041-8205/732/2/L25}.
\adsurl{2011ApJ...732L..25C}.
\end{barticle}
\endbibitem

\bibitem[\protect\citeauthoryear{{Cheng} et~al.}{2023}]{cheng2023deciphering}
\begin{barticle}
\bauthor{\bsnm{{Cheng}}, \binits{X.}},
\bauthor{\bsnm{{Xing}}, \binits{C.}},
\bauthor{\bsnm{{Aulanier}}, \binits{G.}},
\bauthor{\bsnm{{Solanki}}, \binits{S.K.}},
\bauthor{\bsnm{{Peter}}, \binits{H.}},
\bauthor{\bsnm{{Ding}}, \binits{M.D.}}:
\byear{2023},
\batitle{{Deciphering the Slow-rise Precursor of a Major Coronal Mass
  Ejection}}.
\bjtitle{\apjl}
\bvolume{954},
\bfpage{L47}.
\doiurl{https://doi.org/10.3847/2041-8213/acf3e4}.
\adsurl{2023ApJ...954L..47C}.
\end{barticle}
\endbibitem

\bibitem[\protect\citeauthoryear{{Chintzoglou}, {Patsourakos}, and
  {Vourlidas}}{2015}]{chintzoglou2015formation}
\begin{barticle}
\bauthor{\bsnm{{Chintzoglou}}, \binits{G.}},
\bauthor{\bsnm{{Patsourakos}}, \binits{S.}},
\bauthor{\bsnm{{Vourlidas}}, \binits{A.}}:
\byear{2015},
\batitle{{Formation of Magnetic Flux Ropes during a Confined Flaring Well
  before the Onset of a Pair of Major Coronal Mass Ejections}}.
\bjtitle{\apj}
\bvolume{809},
\bfpage{34}.
\doiurl{https://doi.org/10.1088/0004-637X/809/1/34}.
\adsurl{2015ApJ...809...34C}.
\end{barticle}
\endbibitem

\bibitem[\protect\citeauthoryear{{Demoulin} et~al.}{1996}]{demoulin1996quasi}
\begin{barticle}
\bauthor{\bsnm{{Demoulin}}, \binits{P.}},
\bauthor{\bsnm{{Henoux}}, \binits{J.C.}},
\bauthor{\bsnm{{Priest}}, \binits{E.R.}},
\bauthor{\bsnm{{Mandrini}}, \binits{C.H.}}:
\byear{1996},
\batitle{{Quasi-Separatrix layers in solar flares. I. Method.}}
\bjtitle{\aap}
\bvolume{308},
\bfpage{643}.
\adsurl{1996A&A...308..643D}.
\end{barticle}
\endbibitem

\bibitem[\protect\citeauthoryear{{Demoulin} et~al.}{1997}]{demoulin1997quasi}
\begin{barticle}
\bauthor{\bsnm{{Demoulin}}, \binits{P.}},
\bauthor{\bsnm{{Bagala}}, \binits{L.G.}},
\bauthor{\bsnm{{Mandrini}}, \binits{C.H.}},
\bauthor{\bsnm{{Henoux}}, \binits{J.C.}},
\bauthor{\bsnm{{Rovira}}, \binits{M.G.}}:
\byear{1997},
\batitle{{Quasi-separatrix layers in solar flares. II. Observed magnetic
  configurations.}}
\bjtitle{\aap}
\bvolume{325},
\bfpage{305}.
\adsurl{1997A&A...325..305D}.
\end{barticle}
\endbibitem

\bibitem[\protect\citeauthoryear{{Deng} et~al.}{2019}]{deng2019design}
\begin{barticle}
\bauthor{\bsnm{{Deng}}, \binits{Y.-Y.}},
\bauthor{\bsnm{{Zhang}}, \binits{H.-Y.}},
\bauthor{\bsnm{{Yang}}, \binits{J.-F.}},
\bauthor{\bsnm{{Li}}, \binits{F.}},
\bauthor{\bsnm{{Lin}}, \binits{J.-B.}},
\bauthor{\bsnm{{Hou}}, \binits{J.-F.}},
\bauthor{\bsnm{{Wu}}, \binits{Z.}},
\bauthor{\bsnm{{Song}}, \binits{Q.}},
\bauthor{\bsnm{{Duan}}, \binits{W.}},
\bauthor{\bsnm{{Bai}}, \binits{X.-Y.}},
\bauthor{\bsnm{{Wang}}, \binits{D.-G.}},
\bauthor{\bsnm{{Lv}}, \binits{J.}},
\bauthor{\bsnm{{Ge}}, \binits{W.}},
\bauthor{\bsnm{{Wang}}, \binits{J.-N.}},
\bauthor{\bsnm{{Zheng}}, \binits{Z.-Y.}},
\bauthor{\bsnm{{Wang}}, \binits{C.-J.}},
\bauthor{\bsnm{{Wang}}, \binits{N.-G.}},
\bauthor{\bsnm{{Ni}}, \binits{H.-K.}},
\bauthor{\bsnm{{Zeng}}, \binits{Y.-Z.}},
\bauthor{\bsnm{{Zhang}}, \binits{Y.}},
\bauthor{\bsnm{{Yang}}, \binits{X.}},
\bauthor{\bsnm{{Sun}}, \binits{Y.-Z.}},
\bauthor{\bsnm{{Zhang}}, \binits{Z.-Y.}},
\bauthor{\bsnm{{Wang}}, \binits{X.-F.}}:
\byear{2019},
\batitle{{Design of the Full-disk MagnetoGraph (FMG) onboard the ASO-S}}.
\bjtitle{Research in Astronomy and Astrophysics}
\bvolume{19},
\bfpage{157}.
\doiurl{https://doi.org/10.1088/1674-4527/19/11/157}.
\adsurl{2019RAA....19..157D}.
\end{barticle}
\endbibitem

\bibitem[\protect\citeauthoryear{{Dhakal}, {Chintzoglou}, and
  {Zhang}}{2018}]{dhakal2018study}
\begin{barticle}
\bauthor{\bsnm{{Dhakal}}, \binits{S.K.}},
\bauthor{\bsnm{{Chintzoglou}}, \binits{G.}},
\bauthor{\bsnm{{Zhang}}, \binits{J.}}:
\byear{2018},
\batitle{{A Study of a Compound Solar Eruption with Two Consecutive Erupting
  Magnetic Structures}}.
\bjtitle{\apj}
\bvolume{860},
\bfpage{35}.
\doiurl{https://doi.org/10.3847/1538-4357/aac028}.
\adsurl{2018ApJ...860...35D}.
\end{barticle}
\endbibitem

\bibitem[\protect\citeauthoryear{{Ding}, {Hu}, and
  {Wang}}{2006}]{ding2006catastrophic}
\begin{barticle}
\bauthor{\bsnm{{Ding}}, \binits{J.Y.}},
\bauthor{\bsnm{{Hu}}, \binits{Y.Q.}},
\bauthor{\bsnm{{Wang}}, \binits{J.X.}}:
\byear{2006},
\batitle{{Catastrophic Behavior of Multiple Coronal Flux Rope System}}.
\bjtitle{\solphys}
\bvolume{235},
\bfpage{223}.
\doiurl{https://doi.org/10.1007/s11207-006-0092-7}.
\adsurl{2006SoPh..235..223D}.
\end{barticle}
\endbibitem

\bibitem[\protect\citeauthoryear{{Dud{\'\i}k} et~al.}{2014}]{dudik2014slipping}
\begin{barticle}
\bauthor{\bsnm{{Dud{\'\i}k}}, \binits{J.}},
\bauthor{\bsnm{{Janvier}}, \binits{M.}},
\bauthor{\bsnm{{Aulanier}}, \binits{G.}},
\bauthor{\bsnm{{Del Zanna}}, \binits{G.}},
\bauthor{\bsnm{{Karlick{\'y}}}, \binits{M.}},
\bauthor{\bsnm{{Mason}}, \binits{H.E.}},
\bauthor{\bsnm{{Schmieder}}, \binits{B.}}:
\byear{2014},
\batitle{{Slipping Magnetic Reconnection during an X-class Solar Flare Observed
  by SDO/AIA}}.
\bjtitle{\apj}
\bvolume{784},
\bfpage{144}.
\doiurl{https://doi.org/10.1088/0004-637X/784/2/144}.
\adsurl{2014ApJ...784..144D}.
\end{barticle}
\endbibitem

\bibitem[\protect\citeauthoryear{{Dud{\'\i}k} et~al.}{2016}]{dudik2016slipping}
\begin{barticle}
\bauthor{\bsnm{{Dud{\'\i}k}}, \binits{J.}},
\bauthor{\bsnm{{Polito}}, \binits{V.}},
\bauthor{\bsnm{{Janvier}}, \binits{M.}},
\bauthor{\bsnm{{Mulay}}, \binits{S.M.}},
\bauthor{\bsnm{{Karlick{\'y}}}, \binits{M.}},
\bauthor{\bsnm{{Aulanier}}, \binits{G.}},
\bauthor{\bsnm{{Del Zanna}}, \binits{G.}},
\bauthor{\bsnm{{Dzif{\v{c}}{\'a}kov{\'a}}}, \binits{E.}},
\bauthor{\bsnm{{Mason}}, \binits{H.E.}},
\bauthor{\bsnm{{Schmieder}}, \binits{B.}}:
\byear{2016},
\batitle{{Slipping Magnetic Reconnection, Chromospheric Evaporation, Implosion,
  and Precursors in the 2014 September 10 X1.6-Class Solar Flare}}.
\bjtitle{\apj}
\bvolume{823},
\bfpage{41}.
\doiurl{https://doi.org/10.3847/0004-637X/823/1/41}.
\adsurl{2016ApJ...823...41D}.
\end{barticle}
\endbibitem

\bibitem[\protect\citeauthoryear{{Echer}
  et~al.}{2008}]{echer2008interplanetary}
\begin{barticle}
\bauthor{\bsnm{{Echer}}, \binits{E.}},
\bauthor{\bsnm{{Gonzalez}}, \binits{W.D.}},
\bauthor{\bsnm{{Tsurutani}}, \binits{B.T.}},
\bauthor{\bsnm{{Gonzalez}}, \binits{A.L.C.}}:
\byear{2008},
\batitle{{Interplanetary conditions causing intense geomagnetic storms (Dst <=
  -100 nT) during solar cycle 23 (1996-2006)}}.
\bjtitle{Journal of Geophysical Research (Space Physics)}
\bvolume{113},
\bfpage{A05221}.
\doiurl{https://doi.org/10.1029/2007JA012744}.
\adsurl{2008JGRA..113.5221E}.
\end{barticle}
\endbibitem

\bibitem[\protect\citeauthoryear{{Fan}}{2009}]{fan2009emergence}
\begin{barticle}
\bauthor{\bsnm{{Fan}}, \binits{Y.}}:
\byear{2009},
\batitle{{The Emergence of a Twisted Flux Tube into the Solar Atmosphere:
  Sunspot Rotations and the Formation of a Coronal Flux Rope}}.
\bjtitle{\apj}
\bvolume{697},
\bfpage{1529}.
\doiurl{https://doi.org/10.1088/0004-637X/697/2/1529}.
\adsurl{2009ApJ...697.1529F}.
\end{barticle}
\endbibitem

\bibitem[\protect\citeauthoryear{{Farrugia} et~al.}{2006}]{farrugia2006survey}
\begin{barticle}
\bauthor{\bsnm{{Farrugia}}, \binits{C.J.}},
\bauthor{\bsnm{{Matsui}}, \binits{H.}},
\bauthor{\bsnm{{Kucharek}}, \binits{H.}},
\bauthor{\bsnm{{Jordanova}}, \binits{V.K.}},
\bauthor{\bsnm{{Torbert}}, \binits{R.B.}},
\bauthor{\bsnm{{Ogilvie}}, \binits{K.W.}},
\bauthor{\bsnm{{Berdichevsky}}, \binits{D.B.}},
\bauthor{\bsnm{{Smith}}, \binits{C.W.}},
\bauthor{\bsnm{{Skoug}}, \binits{R.}}:
\byear{2006},
\batitle{{Survey of intense Sun Earth connection events (1995 2003)}}.
\bjtitle{Advances in Space Research}
\bvolume{38},
\bfpage{498}.
\doiurl{https://doi.org/10.1016/j.asr.2005.05.051}.
\adsurl{2006AdSpR..38..498F}.
\end{barticle}
\endbibitem

\bibitem[\protect\citeauthoryear{{Gan} et~al.}{2019}]{gan2019advanced}
\begin{barticle}
\bauthor{\bsnm{{Gan}}, \binits{W.-Q.}},
\bauthor{\bsnm{{Zhu}}, \binits{C.}},
\bauthor{\bsnm{{Deng}}, \binits{Y.-Y.}},
\bauthor{\bsnm{{Li}}, \binits{H.}},
\bauthor{\bsnm{{Su}}, \binits{Y.}},
\bauthor{\bsnm{{Zhang}}, \binits{H.-Y.}},
\bauthor{\bsnm{{Chen}}, \binits{B.}},
\bauthor{\bsnm{{Zhang}}, \binits{Z.}},
\bauthor{\bsnm{{Wu}}, \binits{J.}},
\bauthor{\bsnm{{Deng}}, \binits{L.}},
\bauthor{\bsnm{{Huang}}, \binits{Y.}},
\bauthor{\bsnm{{Yang}}, \binits{J.-F.}},
\bauthor{\bsnm{{Cui}}, \binits{J.-J.}},
\bauthor{\bsnm{{Chang}}, \binits{J.}},
\bauthor{\bsnm{{Wang}}, \binits{C.}},
\bauthor{\bsnm{{Wu}}, \binits{J.}},
\bauthor{\bsnm{{Yin}}, \binits{Z.-S.}},
\bauthor{\bsnm{{Chen}}, \binits{W.}},
\bauthor{\bsnm{{Fang}}, \binits{C.}},
\bauthor{\bsnm{{Yan}}, \binits{Y.-H.}},
\bauthor{\bsnm{{Lin}}, \binits{J.}},
\bauthor{\bsnm{{Xiong}}, \binits{W.-M.}},
\bauthor{\bsnm{{Chen}}, \binits{B.}},
\bauthor{\bsnm{{Bao}}, \binits{H.-C.}},
\bauthor{\bsnm{{Cao}}, \binits{C.-X.}},
\bauthor{\bsnm{{Bai}}, \binits{Y.-P.}},
\bauthor{\bsnm{{Wang}}, \binits{T.}},
\bauthor{\bsnm{{Chen}}, \binits{B.-L.}},
\bauthor{\bsnm{{Li}}, \binits{X.-Y.}},
\bauthor{\bsnm{{Zhang}}, \binits{Y.}},
\bauthor{\bsnm{{Feng}}, \binits{L.}},
\bauthor{\bsnm{{Su}}, \binits{J.-T.}},
\bauthor{\bsnm{{Li}}, \binits{Y.}},
\bauthor{\bsnm{{Chen}}, \binits{W.}},
\bauthor{\bsnm{{Li}}, \binits{Y.-P.}},
\bauthor{\bsnm{{Su}}, \binits{Y.-N.}},
\bauthor{\bsnm{{Wu}}, \binits{H.-Y.}},
\bauthor{\bsnm{{Gu}}, \binits{M.}},
\bauthor{\bsnm{{Huang}}, \binits{L.}},
\bauthor{\bsnm{{Tang}}, \binits{X.-J.}}:
\byear{2019},
\batitle{{Advanced Space-based Solar Observatory (ASO-S): an overview}}.
\bjtitle{Research in Astronomy and Astrophysics}
\bvolume{19},
\bfpage{156}.
\doiurl{https://doi.org/10.1088/1674-4527/19/11/156}.
\adsurl{2019RAA....19..156G}.
\end{barticle}
\endbibitem

\bibitem[\protect\citeauthoryear{{Gibson} et~al.}{2006}]{gibson2006evolving}
\begin{barticle}
\bauthor{\bsnm{{Gibson}}, \binits{S.E.}},
\bauthor{\bsnm{{Fan}}, \binits{Y.}},
\bauthor{\bsnm{{T{\"o}r{\"o}k}}, \binits{T.}},
\bauthor{\bsnm{{Kliem}}, \binits{B.}}:
\byear{2006},
\batitle{{The Evolving Sigmoid: Evidence for Magnetic Flux Ropes in the Corona
  Before, During, and After CMES}}.
\bjtitle{\ssr}
\bvolume{124},
\bfpage{131}.
\doiurl{https://doi.org/10.1007/s11214-006-9101-2}.
\adsurl{2006SSRv..124..131G}.
\end{barticle}
\endbibitem

\bibitem[\protect\citeauthoryear{Gonzalez
  et~al.}{1994}]{gonzalez1994geomagnetic}
\begin{barticle}
\bauthor{\bsnm{Gonzalez}, \binits{W.}},
\bauthor{\bsnm{Joselyn}, \binits{J.-A.}},
\bauthor{\bsnm{Kamide}, \binits{Y.}},
\bauthor{\bsnm{Kroehl}, \binits{H.W.}},
\bauthor{\bsnm{Rostoker}, \binits{G.}},
\bauthor{\bsnm{Tsurutani}, \binits{B.T.}},
\bauthor{\bsnm{Vasyliunas}, \binits{V.}}:
\byear{1994},
\batitle{What is a geomagnetic storm?}
\bjtitle{Journal of Geophysical Research: Space Physics}
\bvolume{99},
\bfpage{5771}.
\doiurl{https://doi.org/10.1029/93JA02867}.
\end{barticle}
\endbibitem

\bibitem[\protect\citeauthoryear{{Gou} et~al.}{2019}]{gou2019birth}
\begin{barticle}
\bauthor{\bsnm{{Gou}}, \binits{T.}},
\bauthor{\bsnm{{Liu}}, \binits{R.}},
\bauthor{\bsnm{{Kliem}}, \binits{B.}},
\bauthor{\bsnm{{Wang}}, \binits{Y.}},
\bauthor{\bsnm{{Veronig}}, \binits{A.M.}}:
\byear{2019},
\batitle{{The Birth of A Coronal Mass Ejection}}.
\bjtitle{Science Advances}
\bvolume{5},
\bfpage{7004}.
\doiurl{https://doi.org/10.1126/sciadv.aau7004}.
\adsurl{2019SciA....5.7004G}.
\end{barticle}
\endbibitem

\bibitem[\protect\citeauthoryear{{Green} et~al.}{2018}]{green2018origin}
\begin{barticle}
\bauthor{\bsnm{{Green}}, \binits{L.M.}},
\bauthor{\bsnm{{T{\"o}r{\"o}k}}, \binits{T.}},
\bauthor{\bsnm{{Vr{\v{s}}nak}}, \binits{B.}},
\bauthor{\bsnm{{Manchester}}, \binits{W.}},
\bauthor{\bsnm{{Veronig}}, \binits{A.}}:
\byear{2018},
\batitle{{The Origin, Early Evolution and Predictability of Solar Eruptions}}.
\bjtitle{\ssr}
\bvolume{214},
\bfpage{46}.
\doiurl{https://doi.org/10.1007/s11214-017-0462-5}.
\adsurl{2018SSRv..214...46G}.
\end{barticle}
\endbibitem

\bibitem[\protect\citeauthoryear{{Guo} et~al.}{2010}]{guo2010coexisting}
\begin{barticle}
\bauthor{\bsnm{{Guo}}, \binits{Y.}},
\bauthor{\bsnm{{Schmieder}}, \binits{B.}},
\bauthor{\bsnm{{D{\'e}moulin}}, \binits{P.}},
\bauthor{\bsnm{{Wiegelmann}}, \binits{T.}},
\bauthor{\bsnm{{Aulanier}}, \binits{G.}},
\bauthor{\bsnm{{T{\"o}r{\"o}k}}, \binits{T.}},
\bauthor{\bsnm{{Bommier}}, \binits{V.}}:
\byear{2010},
\batitle{{Coexisting Flux Rope and Dipped Arcade Sections Along One Solar
  Filament}}.
\bjtitle{\apj}
\bvolume{714},
\bfpage{343}.
\doiurl{https://doi.org/10.1088/0004-637X/714/1/343}.
\adsurl{2010ApJ...714..343G}.
\end{barticle}
\endbibitem

\bibitem[\protect\citeauthoryear{Hale and Nicholson}{1938}]{hale1938magnetic}
\begin{botherref}
\oauthor{\bsnm{Hale}, \binits{G.E.}},
\oauthor{\bsnm{Nicholson}, \binits{S.B.}}:
1938,
Magnetic observations of sunspots, 1917-1924...
\textit{Washington}.
\end{botherref}
\endbibitem

\bibitem[\protect\citeauthoryear{{Harten} and {Clark}}{1995}]{harten1995design}
\begin{barticle}
\bauthor{\bsnm{{Harten}}, \binits{R.}},
\bauthor{\bsnm{{Clark}}, \binits{K.}}:
\byear{1995},
\batitle{{The Design Features of the GGS Wind and Polar Spacecraft}}.
\bjtitle{\ssr}
\bvolume{71},
\bfpage{23}.
\doiurl{https://doi.org/10.1007/BF00751324}.
\adsurl{1995SSRv...71...23H}.
\end{barticle}
\endbibitem

\bibitem[\protect\citeauthoryear{{Hou} et~al.}{2018}]{hou2018eruption}
\begin{barticle}
\bauthor{\bsnm{{Hou}}, \binits{Y.J.}},
\bauthor{\bsnm{{Zhang}}, \binits{J.}},
\bauthor{\bsnm{{Li}}, \binits{T.}},
\bauthor{\bsnm{{Yang}}, \binits{S.H.}},
\bauthor{\bsnm{{Li}}, \binits{X.H.}}:
\byear{2018},
\batitle{{Eruption of a multi-flux-rope system in solar active region 12673
  leading to the two largest flares in Solar Cycle 24}}.
\bjtitle{\aap}
\bvolume{619},
\bfpage{A100}.
\doiurl{https://doi.org/10.1051/0004-6361/201732530}.
\adsurl{2018A&A...619A.100H}.
\end{barticle}
\endbibitem

\bibitem[\protect\citeauthoryear{{Hou} et~al.}{2023}]{hou2023partial}
\begin{barticle}
\bauthor{\bsnm{{Hou}}, \binits{Y.}},
\bauthor{\bsnm{{Li}}, \binits{C.}},
\bauthor{\bsnm{{Li}}, \binits{T.}},
\bauthor{\bsnm{{Su}}, \binits{J.}},
\bauthor{\bsnm{{Qiu}}, \binits{Y.}},
\bauthor{\bsnm{{Yang}}, \binits{S.}},
\bauthor{\bsnm{{Yang}}, \binits{L.}},
\bauthor{\bsnm{{Li}}, \binits{L.}},
\bauthor{\bsnm{{Guo}}, \binits{Y.}},
\bauthor{\bsnm{{Hou}}, \binits{Z.}},
\bauthor{\bsnm{{Song}}, \binits{Q.}},
\bauthor{\bsnm{{Bai}}, \binits{X.}},
\bauthor{\bsnm{{Zhou}}, \binits{G.}},
\bauthor{\bsnm{{Ding}}, \binits{M.}},
\bauthor{\bsnm{{Gan}}, \binits{W.}},
\bauthor{\bsnm{{Deng}}, \binits{Y.}}:
\byear{2023},
\batitle{{Partial Eruption of Solar Filaments. I. Configuration and Formation
  of Double-decker Filaments}}.
\bjtitle{\apj}
\bvolume{959},
\bfpage{69}.
\doiurl{https://doi.org/10.3847/1538-4357/ad08bd}.
\adsurl{2023ApJ...959...69H}.
\end{barticle}
\endbibitem

\bibitem[\protect\citeauthoryear{{Inoue} and {Kusano}}{2006}]{inoue2006three}
\begin{barticle}
\bauthor{\bsnm{{Inoue}}, \binits{S.}},
\bauthor{\bsnm{{Kusano}}, \binits{K.}}:
\byear{2006},
\batitle{{Three-dimensional Simulation Study of Flux Rope Dynamics in the Solar
  Corona}}.
\bjtitle{\apj}
\bvolume{645},
\bfpage{742}.
\doiurl{https://doi.org/10.1086/503153}.
\adsurl{2006ApJ...645..742I}.
\end{barticle}
\endbibitem

\bibitem[\protect\citeauthoryear{{Janvier} et~al.}{2013}]{janvier2013standard}
\begin{barticle}
\bauthor{\bsnm{{Janvier}}, \binits{M.}},
\bauthor{\bsnm{{Aulanier}}, \binits{G.}},
\bauthor{\bsnm{{Pariat}}, \binits{E.}},
\bauthor{\bsnm{{D{\'e}moulin}}, \binits{P.}}:
\byear{2013},
\batitle{{The standard flare model in three dimensions. III. Slip-running
  reconnection properties}}.
\bjtitle{\aap}
\bvolume{555},
\bfpage{A77}.
\doiurl{https://doi.org/10.1051/0004-6361/201321164}.
\adsurl{2013A&A...555A..77J}.
\end{barticle}
\endbibitem

\bibitem[\protect\citeauthoryear{{Jiang}
  et~al.}{2013}]{jiang2013magnetohydrodynamic}
\begin{barticle}
\bauthor{\bsnm{{Jiang}}, \binits{C.}},
\bauthor{\bsnm{{Feng}}, \binits{X.}},
\bauthor{\bsnm{{Wu}}, \binits{S.T.}},
\bauthor{\bsnm{{Hu}}, \binits{Q.}}:
\byear{2013},
\batitle{{Magnetohydrodynamic Simulation of a Sigmoid Eruption of Active Region
  11283}}.
\bjtitle{\apjl}
\bvolume{771},
\bfpage{L30}.
\doiurl{https://doi.org/10.1088/2041-8205/771/2/L30}.
\adsurl{2013ApJ...771L..30J}.
\end{barticle}
\endbibitem

\bibitem[\protect\citeauthoryear{{Jiang}
  et~al.}{2018}]{jiang2018magnetohydrodynamic}
\begin{barticle}
\bauthor{\bsnm{{Jiang}}, \binits{C.}},
\bauthor{\bsnm{{Zou}}, \binits{P.}},
\bauthor{\bsnm{{Feng}}, \binits{X.}},
\bauthor{\bsnm{{Hu}}, \binits{Q.}},
\bauthor{\bsnm{{Liu}}, \binits{R.}},
\bauthor{\bsnm{{Vemareddy}}, \binits{P.}},
\bauthor{\bsnm{{Duan}}, \binits{A.}},
\bauthor{\bsnm{{Zuo}}, \binits{P.}},
\bauthor{\bsnm{{Wang}}, \binits{Y.}},
\bauthor{\bsnm{{Wei}}, \binits{F.}}:
\byear{2018},
\batitle{{Magnetohydrodynamic Simulation of the X9.3 Flare on 2017 September 6:
  Evolving Magnetic Topology}}.
\bjtitle{\apj}
\bvolume{869},
\bfpage{13}.
\doiurl{https://doi.org/10.3847/1538-4357/aaeacc}.
\adsurl{2018ApJ...869...13J}.
\end{barticle}
\endbibitem

\bibitem[\protect\citeauthoryear{{Jiang}
  et~al.}{2019}]{jiang2019reconstruction}
\begin{barticle}
\bauthor{\bsnm{{Jiang}}, \binits{C.}},
\bauthor{\bsnm{{Duan}}, \binits{A.}},
\bauthor{\bsnm{{Feng}}, \binits{X.}},
\bauthor{\bsnm{{Zou}}, \binits{P.}},
\bauthor{\bsnm{{Zuo}}, \binits{P.}},
\bauthor{\bsnm{{Wang}}, \binits{Y.}}:
\byear{2019},
\batitle{{Reconstruction of a Highly Twisted Magnetic Flux Rope for an
  Inter-Active-Region X-Class Solar Flare}}.
\bjtitle{Frontiers in Astronomy and Space Sciences}
\bvolume{6},
\bfpage{63}.
\doiurl{https://doi.org/10.3389/fspas.2019.00063}.
\adsurl{2019FrASS...6...63J}.
\end{barticle}
\endbibitem

\bibitem[\protect\citeauthoryear{{Joshi} et~al.}{2018}]{joshi2018major}
\begin{barticle}
\bauthor{\bsnm{{Joshi}}, \binits{B.}},
\bauthor{\bsnm{{Ibrahim}}, \binits{M.S.}},
\bauthor{\bsnm{{Shanmugaraju}}, \binits{A.}},
\bauthor{\bsnm{{Chakrabarty}}, \binits{D.}}:
\byear{2018},
\batitle{{A Major Geoeffective CME from NOAA 12371: Initiation, CME-CME
  Interactions, and Interplanetary Consequences}}.
\bjtitle{\solphys}
\bvolume{293},
\bfpage{107}.
\doiurl{https://doi.org/10.1007/s11207-018-1325-2}.
\adsurl{2018SoPh..293..107J}.
\end{barticle}
\endbibitem

\bibitem[\protect\citeauthoryear{{Kliem} et~al.}{2014}]{kliem2014slow}
\begin{barticle}
\bauthor{\bsnm{{Kliem}}, \binits{B.}},
\bauthor{\bsnm{{T{\"o}r{\"o}k}}, \binits{T.}},
\bauthor{\bsnm{{Titov}}, \binits{V.S.}},
\bauthor{\bsnm{{Lionello}}, \binits{R.}},
\bauthor{\bsnm{{Linker}}, \binits{J.A.}},
\bauthor{\bsnm{{Liu}}, \binits{R.}},
\bauthor{\bsnm{{Liu}}, \binits{C.}},
\bauthor{\bsnm{{Wang}}, \binits{H.}}:
\byear{2014},
\batitle{{Slow Rise and Partial Eruption of a Double-decker Filament. II. A
  Double Flux Rope Model}}.
\bjtitle{\apj}
\bvolume{792},
\bfpage{107}.
\doiurl{https://doi.org/10.1088/0004-637X/792/2/107}.
\adsurl{2014ApJ...792..107K}.
\end{barticle}
\endbibitem

\bibitem[\protect\citeauthoryear{{Krall} et~al.}{2001}]{krall2001erupting}
\begin{barticle}
\bauthor{\bsnm{{Krall}}, \binits{J.}},
\bauthor{\bsnm{{Chen}}, \binits{J.}},
\bauthor{\bsnm{{Duffin}}, \binits{R.T.}},
\bauthor{\bsnm{{Howard}}, \binits{R.A.}},
\bauthor{\bsnm{{Thompson}}, \binits{B.J.}}:
\byear{2001},
\batitle{{Erupting Solar Magnetic Flux Ropes: Theory and Observation}}.
\bjtitle{\apj}
\bvolume{562},
\bfpage{1045}.
\doiurl{https://doi.org/10.1086/323844}.
\adsurl{2001ApJ...562.1045K}.
\end{barticle}
\endbibitem

\bibitem[\protect\citeauthoryear{{Lakhina} and
  {Tsurutani}}{2016}]{lakhina2016geomagnetic}
\begin{barticle}
\bauthor{\bsnm{{Lakhina}}, \binits{G.S.}},
\bauthor{\bsnm{{Tsurutani}}, \binits{B.T.}}:
\byear{2016},
\batitle{{Geomagnetic storms: historical perspective to modern view}}.
\bjtitle{Geoscience Letters}
\bvolume{3},
\bfpage{5}.
\doiurl{https://doi.org/10.1186/s40562-016-0037-4}.
\adsurl{2016GSL.....3....5L}.
\end{barticle}
\endbibitem

\bibitem[\protect\citeauthoryear{{Leka} et~al.}{1996}]{leka1996evidence}
\begin{barticle}
\bauthor{\bsnm{{Leka}}, \binits{K.D.}},
\bauthor{\bsnm{{Canfield}}, \binits{R.C.}},
\bauthor{\bsnm{{McClymont}}, \binits{A.N.}},
\bauthor{\bsnm{{van Driel-Gesztelyi}}, \binits{L.}}:
\byear{1996},
\batitle{{Evidence for Current-carrying Emerging Flux}}.
\bjtitle{\apj}
\bvolume{462},
\bfpage{547}.
\doiurl{https://doi.org/10.1086/177171}.
\adsurl{1996ApJ...462..547L}.
\end{barticle}
\endbibitem

\bibitem[\protect\citeauthoryear{{Lemen} et~al.}{2012}]{lemen2012atmospheric}
\begin{barticle}
\bauthor{\bsnm{{Lemen}}, \binits{J.R.}},
\bauthor{\bsnm{{Title}}, \binits{A.M.}},
\bauthor{\bsnm{{Akin}}, \binits{D.J.}},
\bauthor{\bsnm{{Boerner}}, \binits{P.F.}},
\bauthor{\bsnm{{Chou}}, \binits{C.}},
\bauthor{\bsnm{{Drake}}, \binits{J.F.}},
\bauthor{\bsnm{{Duncan}}, \binits{D.W.}},
\bauthor{\bsnm{{Edwards}}, \binits{C.G.}},
\bauthor{\bsnm{{Friedlaender}}, \binits{F.M.}},
\bauthor{\bsnm{{Heyman}}, \binits{G.F.}},
\bauthor{\bsnm{{Hurlburt}}, \binits{N.E.}},
\bauthor{\bsnm{{Katz}}, \binits{N.L.}},
\bauthor{\bsnm{{Kushner}}, \binits{G.D.}},
\bauthor{\bsnm{{Levay}}, \binits{M.}},
\bauthor{\bsnm{{Lindgren}}, \binits{R.W.}},
\bauthor{\bsnm{{Mathur}}, \binits{D.P.}},
\bauthor{\bsnm{{McFeaters}}, \binits{E.L.}},
\bauthor{\bsnm{{Mitchell}}, \binits{S.}},
\bauthor{\bsnm{{Rehse}}, \binits{R.A.}},
\bauthor{\bsnm{{Schrijver}}, \binits{C.J.}},
\bauthor{\bsnm{{Springer}}, \binits{L.A.}},
\bauthor{\bsnm{{Stern}}, \binits{R.A.}},
\bauthor{\bsnm{{Tarbell}}, \binits{T.D.}},
\bauthor{\bsnm{{Wuelser}}, \binits{J.-P.}},
\bauthor{\bsnm{{Wolfson}}, \binits{C.J.}},
\bauthor{\bsnm{{Yanari}}, \binits{C.}},
\bauthor{\bsnm{{Bookbinder}}, \binits{J.A.}},
\bauthor{\bsnm{{Cheimets}}, \binits{P.N.}},
\bauthor{\bsnm{{Caldwell}}, \binits{D.}},
\bauthor{\bsnm{{Deluca}}, \binits{E.E.}},
\bauthor{\bsnm{{Gates}}, \binits{R.}},
\bauthor{\bsnm{{Golub}}, \binits{L.}},
\bauthor{\bsnm{{Park}}, \binits{S.}},
\bauthor{\bsnm{{Podgorski}}, \binits{W.A.}},
\bauthor{\bsnm{{Bush}}, \binits{R.I.}},
\bauthor{\bsnm{{Scherrer}}, \binits{P.H.}},
\bauthor{\bsnm{{Gummin}}, \binits{M.A.}},
\bauthor{\bsnm{{Smith}}, \binits{P.}},
\bauthor{\bsnm{{Auker}}, \binits{G.}},
\bauthor{\bsnm{{Jerram}}, \binits{P.}},
\bauthor{\bsnm{{Pool}}, \binits{P.}},
\bauthor{\bsnm{{Soufli}}, \binits{R.}},
\bauthor{\bsnm{{Windt}}, \binits{D.L.}},
\bauthor{\bsnm{{Beardsley}}, \binits{S.}},
\bauthor{\bsnm{{Clapp}}, \binits{M.}},
\bauthor{\bsnm{{Lang}}, \binits{J.}},
\bauthor{\bsnm{{Waltham}}, \binits{N.}}:
\byear{2012},
\batitle{{The Atmospheric Imaging Assembly (AIA) on the Solar Dynamics
  Observatory (SDO)}}.
\bjtitle{\solphys}
\bvolume{275},
\bfpage{17}.
\doiurl{https://doi.org/10.1007/s11207-011-9776-8}.
\adsurl{2012SoPh..275...17L}.
\end{barticle}
\endbibitem

\bibitem[\protect\citeauthoryear{{Li} et~al.}{2022}]{li2022chinese}
\begin{barticle}
\bauthor{\bsnm{{Li}}, \binits{C.}},
\bauthor{\bsnm{{Fang}}, \binits{C.}},
\bauthor{\bsnm{{Li}}, \binits{Z.}},
\bauthor{\bsnm{{Ding}}, \binits{M.}},
\bauthor{\bsnm{{Chen}}, \binits{P.}},
\bauthor{\bsnm{{Qiu}}, \binits{Y.}},
\bauthor{\bsnm{{You}}, \binits{W.}},
\bauthor{\bsnm{{Yuan}}, \binits{Y.}},
\bauthor{\bsnm{{An}}, \binits{M.}},
\bauthor{\bsnm{{Tao}}, \binits{H.}},
\bauthor{\bsnm{{Li}}, \binits{X.}},
\bauthor{\bsnm{{Chen}}, \binits{Z.}},
\bauthor{\bsnm{{Liu}}, \binits{Q.}},
\bauthor{\bsnm{{Mei}}, \binits{G.}},
\bauthor{\bsnm{{Yang}}, \binits{L.}},
\bauthor{\bsnm{{Zhang}}, \binits{W.}},
\bauthor{\bsnm{{Cheng}}, \binits{W.}},
\bauthor{\bsnm{{Chen}}, \binits{J.}},
\bauthor{\bsnm{{Chen}}, \binits{C.}},
\bauthor{\bsnm{{Gu}}, \binits{Q.}},
\bauthor{\bsnm{{Huang}}, \binits{Q.}},
\bauthor{\bsnm{{Liu}}, \binits{M.}},
\bauthor{\bsnm{{Han}}, \binits{C.}},
\bauthor{\bsnm{{Xin}}, \binits{H.}},
\bauthor{\bsnm{{Chen}}, \binits{C.}},
\bauthor{\bsnm{{Ni}}, \binits{Y.}},
\bauthor{\bsnm{{Wang}}, \binits{W.}},
\bauthor{\bsnm{{Rao}}, \binits{S.}},
\bauthor{\bsnm{{Li}}, \binits{H.}},
\bauthor{\bsnm{{Lu}}, \binits{X.}},
\bauthor{\bsnm{{Wang}}, \binits{W.}},
\bauthor{\bsnm{{Lin}}, \binits{J.}},
\bauthor{\bsnm{{Jiang}}, \binits{Y.}},
\bauthor{\bsnm{{Meng}}, \binits{L.}},
\bauthor{\bsnm{{Zhao}}, \binits{J.}}:
\byear{2022},
\batitle{{The Chinese H{\ensuremath{\alpha}} Solar Explorer (CHASE) mission: An
  overview}}.
\bjtitle{Science China Physics, Mechanics, and Astronomy}
\bvolume{65},
\bfpage{289602}.
\doiurl{https://doi.org/10.1007/s11433-022-1893-3}.
\adsurl{2022SCPMA..6589602L}.
\end{barticle}
\endbibitem

\bibitem[\protect\citeauthoryear{{Li} et~al.}{2012}]{li2012twin}
\begin{barticle}
\bauthor{\bsnm{{Li}}, \binits{G.}},
\bauthor{\bsnm{{Moore}}, \binits{R.}},
\bauthor{\bsnm{{Mewaldt}}, \binits{R.A.}},
\bauthor{\bsnm{{Zhao}}, \binits{L.}},
\bauthor{\bsnm{{Labrador}}, \binits{A.W.}}:
\byear{2012},
\batitle{{A Twin-CME Scenario for Ground Level Enhancement Events}}.
\bjtitle{\ssr}
\bvolume{171},
\bfpage{141}.
\doiurl{https://doi.org/10.1007/s11214-011-9823-7}.
\adsurl{2012SSRv..171..141L}.
\end{barticle}
\endbibitem

\bibitem[\protect\citeauthoryear{{Li} et~al.}{2019}]{li2019lyman}
\begin{barticle}
\bauthor{\bsnm{{Li}}, \binits{H.}},
\bauthor{\bsnm{{Chen}}, \binits{B.}},
\bauthor{\bsnm{{Feng}}, \binits{L.}},
\bauthor{\bsnm{{Li}}, \binits{Y.}},
\bauthor{\bsnm{{Huang}}, \binits{Y.}},
\bauthor{\bsnm{{Li}}, \binits{J.-W.}},
\bauthor{\bsnm{{Lu}}, \binits{L.}},
\bauthor{\bsnm{{Xue}}, \binits{J.-C.}},
\bauthor{\bsnm{{Ying}}, \binits{B.-L.}},
\bauthor{\bsnm{{Zhao}}, \binits{J.}},
\bauthor{\bsnm{{Yang}}, \binits{Y.-T.}},
\bauthor{\bsnm{{Gan}}, \binits{W.-Q.}},
\bauthor{\bsnm{{Fang}}, \binits{C.}},
\bauthor{\bsnm{{Song}}, \binits{K.-F.}},
\bauthor{\bsnm{{Wang}}, \binits{H.}},
\bauthor{\bsnm{{Guo}}, \binits{Q.-F.}},
\bauthor{\bsnm{{He}}, \binits{L.-P.}},
\bauthor{\bsnm{{Zhu}}, \binits{B.}},
\bauthor{\bsnm{{Zhu}}, \binits{C.}},
\bauthor{\bsnm{{Deng}}, \binits{L.}},
\bauthor{\bsnm{{Bao}}, \binits{H.-C.}},
\bauthor{\bsnm{{Cao}}, \binits{C.-X.}},
\bauthor{\bsnm{{Yang}}, \binits{Z.-G.}}:
\byear{2019},
\batitle{{The Lyman-alpha Solar Telescope (LST) for the ASO-S mission
  {\textemdash} I. Scientific objectives and overview}}.
\bjtitle{Research in Astronomy and Astrophysics}
\bvolume{19},
\bfpage{158}.
\doiurl{https://doi.org/10.1088/1674-4527/19/11/158}.
\adsurl{2019RAA....19..158L}.
\end{barticle}
\endbibitem

\bibitem[\protect\citeauthoryear{{Li} and {Zhang}}{2014}]{li2014slipping}
\begin{barticle}
\bauthor{\bsnm{{Li}}, \binits{T.}},
\bauthor{\bsnm{{Zhang}}, \binits{J.}}:
\byear{2014},
\batitle{{Slipping Magnetic Reconnection Triggering a Solar Eruption of a
  Triangle-shaped Flag Flux Rope}}.
\bjtitle{\apjl}
\bvolume{791},
\bfpage{L13}.
\doiurl{https://doi.org/10.1088/2041-8205/791/1/L13}.
\adsurl{2014ApJ...791L..13L}.
\end{barticle}
\endbibitem

\bibitem[\protect\citeauthoryear{{Li}, {Priest}, and {Guo}}{2021}]{li2021three}
\begin{barticle}
\bauthor{\bsnm{{Li}}, \binits{T.}},
\bauthor{\bsnm{{Priest}}, \binits{E.}},
\bauthor{\bsnm{{Guo}}, \binits{R.}}:
\byear{2021},
\batitle{{Three-dimensional magnetic reconnection in astrophysical plasmas}}.
\bjtitle{Proceedings of the Royal Society of London Series A}
\bvolume{477},
\bfpage{20200949}.
\doiurl{https://doi.org/10.1098/rspa.2020.0949}.
\adsurl{2021RSPSA.47700949L}.
\end{barticle}
\endbibitem

\bibitem[\protect\citeauthoryear{{Li} et~al.}{2016}]{li2016slipping}
\begin{barticle}
\bauthor{\bsnm{{Li}}, \binits{T.}},
\bauthor{\bsnm{{Yang}}, \binits{K.}},
\bauthor{\bsnm{{Hou}}, \binits{Y.}},
\bauthor{\bsnm{{Zhang}}, \binits{J.}}:
\byear{2016},
\batitle{{Slipping Magnetic Reconnection of Flux-rope Structures as a Precursor
  to an Eruptive X-class Solar Flare}}.
\bjtitle{\apj}
\bvolume{830},
\bfpage{152}.
\doiurl{https://doi.org/10.3847/0004-637X/830/2/152}.
\adsurl{2016ApJ...830..152L}.
\end{barticle}
\endbibitem

\bibitem[\protect\citeauthoryear{{Li} et~al.}{2018a}]{li2018three}
\begin{barticle}
\bauthor{\bsnm{{Li}}, \binits{T.}},
\bauthor{\bsnm{{Hou}}, \binits{Y.}},
\bauthor{\bsnm{{Yang}}, \binits{S.}},
\bauthor{\bsnm{{Zhang}}, \binits{J.}}:
\byear{2018}a,
\batitle{{Three-dimensional Magnetic Reconnection Triggering an X-class
  Confined Flare in Active Region 12192}}.
\bjtitle{\apj}
\bvolume{869},
\bfpage{172}.
\doiurl{https://doi.org/10.3847/1538-4357/aaefee}.
\adsurl{2018ApJ...869..172L}.
\end{barticle}
\endbibitem

\bibitem[\protect\citeauthoryear{{Li} et~al.}{2018b}]{li2018two}
\begin{barticle}
\bauthor{\bsnm{{Li}}, \binits{T.}},
\bauthor{\bsnm{{Yang}}, \binits{S.}},
\bauthor{\bsnm{{Zhang}}, \binits{Q.}},
\bauthor{\bsnm{{Hou}}, \binits{Y.}},
\bauthor{\bsnm{{Zhang}}, \binits{J.}}:
\byear{2018}b,
\batitle{{Two Episodes of Magnetic Reconnections during a Confined
  Circular-ribbon Flare}}.
\bjtitle{\apj}
\bvolume{859},
\bfpage{122}.
\doiurl{https://doi.org/10.3847/1538-4357/aabe84}.
\adsurl{2018ApJ...859..122L}.
\end{barticle}
\endbibitem

\bibitem[\protect\citeauthoryear{{Liu} et~al.}{2018}]{liu2018rapid}
\begin{barticle}
\bauthor{\bsnm{{Liu}}, \binits{L.}},
\bauthor{\bsnm{{Cheng}}, \binits{X.}},
\bauthor{\bsnm{{Wang}}, \binits{Y.}},
\bauthor{\bsnm{{Zhou}}, \binits{Z.}},
\bauthor{\bsnm{{Guo}}, \binits{Y.}},
\bauthor{\bsnm{{Cui}}, \binits{J.}}:
\byear{2018},
\batitle{{Rapid Buildup of a Magnetic Flux Rope during a Confined X2.2 Class
  Flare in NOAA AR 12673}}.
\bjtitle{\apjl}
\bvolume{867},
\bfpage{L5}.
\doiurl{https://doi.org/10.3847/2041-8213/aae826}.
\adsurl{2018ApJ...867L...5L}.
\end{barticle}
\endbibitem

\bibitem[\protect\citeauthoryear{{Liu} et~al.}{2019}]{liu2019formation}
\begin{barticle}
\bauthor{\bsnm{{Liu}}, \binits{L.}},
\bauthor{\bsnm{{Cheng}}, \binits{X.}},
\bauthor{\bsnm{{Wang}}, \binits{Y.}},
\bauthor{\bsnm{{Zhou}}, \binits{Z.}}:
\byear{2019},
\batitle{{Formation of a Magnetic Flux Rope in the Early Emergence Phase of
  NOAA Active Region 12673}}.
\bjtitle{\apj}
\bvolume{884},
\bfpage{45}.
\doiurl{https://doi.org/10.3847/1538-4357/ab3c6c}.
\adsurl{2019ApJ...884...45L}.
\end{barticle}
\endbibitem

\bibitem[\protect\citeauthoryear{{Liu}}{2020}]{liu2020magnetic}
\begin{barticle}
\bauthor{\bsnm{{Liu}}, \binits{R.}}:
\byear{2020},
\batitle{{Magnetic flux ropes in the solar corona: structure and evolution
  toward eruption}}.
\bjtitle{Research in Astronomy and Astrophysics}
\bvolume{20},
\bfpage{165}.
\doiurl{https://doi.org/10.1088/1674-4527/20/10/165}.
\adsurl{2020RAA....20..165L}.
\end{barticle}
\endbibitem

\bibitem[\protect\citeauthoryear{{Liu} et~al.}{2010}]{liu2010sigmoid}
\begin{barticle}
\bauthor{\bsnm{{Liu}}, \binits{R.}},
\bauthor{\bsnm{{Liu}}, \binits{C.}},
\bauthor{\bsnm{{Wang}}, \binits{S.}},
\bauthor{\bsnm{{Deng}}, \binits{N.}},
\bauthor{\bsnm{{Wang}}, \binits{H.}}:
\byear{2010},
\batitle{{Sigmoid-to-flux-rope Transition Leading to a Loop-like Coronal Mass
  Ejection}}.
\bjtitle{\apjl}
\bvolume{725},
\bfpage{L84}.
\doiurl{https://doi.org/10.1088/2041-8205/725/1/L84}.
\adsurl{2010ApJ...725L..84L}.
\end{barticle}
\endbibitem

\bibitem[\protect\citeauthoryear{{Liu} et~al.}{2016}]{liu2016structure}
\begin{barticle}
\bauthor{\bsnm{{Liu}}, \binits{R.}},
\bauthor{\bsnm{{Kliem}}, \binits{B.}},
\bauthor{\bsnm{{Titov}}, \binits{V.S.}},
\bauthor{\bsnm{{Chen}}, \binits{J.}},
\bauthor{\bsnm{{Wang}}, \binits{Y.}},
\bauthor{\bsnm{{Wang}}, \binits{H.}},
\bauthor{\bsnm{{Liu}}, \binits{C.}},
\bauthor{\bsnm{{Xu}}, \binits{Y.}},
\bauthor{\bsnm{{Wiegelmann}}, \binits{T.}}:
\byear{2016},
\batitle{{Structure, Stability, and Evolution of Magnetic Flux Ropes from the
  Perspective of Magnetic Twist}}.
\bjtitle{\apj}
\bvolume{818},
\bfpage{148}.
\doiurl{https://doi.org/10.3847/0004-637X/818/2/148}.
\adsurl{2016ApJ...818..148L}.
\end{barticle}
\endbibitem

\bibitem[\protect\citeauthoryear{{Liu} et~al.}{2014a}]{liu2014observations}
\begin{barticle}
\bauthor{\bsnm{{Liu}}, \binits{Y.D.}},
\bauthor{\bsnm{{Luhmann}}, \binits{J.G.}},
\bauthor{\bsnm{{Kajdi{\v{c}}}}, \binits{P.}},
\bauthor{\bsnm{{Kilpua}}, \binits{E.K.J.}},
\bauthor{\bsnm{{Lugaz}}, \binits{N.}},
\bauthor{\bsnm{{Nitta}}, \binits{N.V.}},
\bauthor{\bsnm{{M{\"o}stl}}, \binits{C.}},
\bauthor{\bsnm{{Lavraud}}, \binits{B.}},
\bauthor{\bsnm{{Bale}}, \binits{S.D.}},
\bauthor{\bsnm{{Farrugia}}, \binits{C.J.}},
\bauthor{\bsnm{{Galvin}}, \binits{A.B.}}:
\byear{2014}a,
\batitle{{Observations of an extreme storm in interplanetary space caused by
  successive coronal mass ejections}}.
\bjtitle{Nature Communications}
\bvolume{5},
\bfpage{3481}.
\doiurl{https://doi.org/10.1038/ncomms4481}.
\adsurl{2014NatCo...5.3481L}.
\end{barticle}
\endbibitem

\bibitem[\protect\citeauthoryear{{Liu} et~al.}{2014b}]{liu2014new}
\begin{barticle}
\bauthor{\bsnm{{Liu}}, \binits{Z.}},
\bauthor{\bsnm{{Xu}}, \binits{J.}},
\bauthor{\bsnm{{Gu}}, \binits{B.-Z.}},
\bauthor{\bsnm{{Wang}}, \binits{S.}},
\bauthor{\bsnm{{You}}, \binits{J.-Q.}},
\bauthor{\bsnm{{Shen}}, \binits{L.-X.}},
\bauthor{\bsnm{{Lu}}, \binits{R.-W.}},
\bauthor{\bsnm{{Jin}}, \binits{Z.-Y.}},
\bauthor{\bsnm{{Chen}}, \binits{L.-F.}},
\bauthor{\bsnm{{Lou}}, \binits{K.}},
\bauthor{\bsnm{{Li}}, \binits{Z.}},
\bauthor{\bsnm{{Liu}}, \binits{G.-Q.}},
\bauthor{\bsnm{{Xu}}, \binits{Z.}},
\bauthor{\bsnm{{Rao}}, \binits{C.-H.}},
\bauthor{\bsnm{{Hu}}, \binits{Q.-Q.}},
\bauthor{\bsnm{{Li}}, \binits{R.-F.}},
\bauthor{\bsnm{{Fu}}, \binits{H.-W.}},
\bauthor{\bsnm{{Wang}}, \binits{F.}},
\bauthor{\bsnm{{Bao}}, \binits{M.-X.}},
\bauthor{\bsnm{{Wu}}, \binits{M.-C.}},
\bauthor{\bsnm{{Zhang}}, \binits{B.-R.}}:
\byear{2014}b,
\batitle{{New vacuum solar telescope and observations with high resolution}}.
\bjtitle{Research in Astronomy and Astrophysics}
\bvolume{14},
\bfpage{705}.
\doiurl{https://doi.org/10.1088/1674-4527/14/6/009}.
\adsurl{2014RAA....14..705L}.
\end{barticle}
\endbibitem

\bibitem[\protect\citeauthoryear{{Lugaz} et~al.}{2017}]{lugaz2017interaction}
\begin{barticle}
\bauthor{\bsnm{{Lugaz}}, \binits{N.}},
\bauthor{\bsnm{{Temmer}}, \binits{M.}},
\bauthor{\bsnm{{Wang}}, \binits{Y.}},
\bauthor{\bsnm{{Farrugia}}, \binits{C.J.}}:
\byear{2017},
\batitle{{The Interaction of Successive Coronal Mass Ejections: A Review}}.
\bjtitle{\solphys}
\bvolume{292},
\bfpage{64}.
\doiurl{https://doi.org/10.1007/s11207-017-1091-6}.
\adsurl{2017SoPh..292...64L}.
\end{barticle}
\endbibitem

\bibitem[\protect\citeauthoryear{{Manchester}
  et~al.}{2004}]{manchester2004eruption}
\begin{barticle}
\bauthor{\bsnm{{Manchester}}, \binits{I.} \bsuffix{W.}},
\bauthor{\bsnm{{Gombosi}}, \binits{T.}},
\bauthor{\bsnm{{DeZeeuw}}, \binits{D.}},
\bauthor{\bsnm{{Fan}}, \binits{Y.}}:
\byear{2004},
\batitle{{Eruption of a Buoyantly Emerging Magnetic Flux Rope}}.
\bjtitle{\apj}
\bvolume{610},
\bfpage{588}.
\doiurl{https://doi.org/10.1086/421516}.
\adsurl{2004ApJ...610..588M}.
\end{barticle}
\endbibitem

\bibitem[\protect\citeauthoryear{{Nindos} et~al.}{2015}]{nindos2015common}
\begin{barticle}
\bauthor{\bsnm{{Nindos}}, \binits{A.}},
\bauthor{\bsnm{{Patsourakos}}, \binits{S.}},
\bauthor{\bsnm{{Vourlidas}}, \binits{A.}},
\bauthor{\bsnm{{Tagikas}}, \binits{C.}}:
\byear{2015},
\batitle{{How Common Are Hot Magnetic Flux Ropes in the Low Solar Corona? A
  Statistical Study of EUV Observations}}.
\bjtitle{\apj}
\bvolume{808},
\bfpage{117}.
\doiurl{https://doi.org/10.1088/0004-637X/808/2/117}.
\adsurl{2015ApJ...808..117N}.
\end{barticle}
\endbibitem

\bibitem[\protect\citeauthoryear{{Patsourakos} and
  {Vourlidas}}{2012}]{patsourakos2012nature}
\begin{barticle}
\bauthor{\bsnm{{Patsourakos}}, \binits{S.}},
\bauthor{\bsnm{{Vourlidas}}, \binits{A.}}:
\byear{2012},
\batitle{{On the Nature and Genesis of EUV Waves: A Synthesis of Observations
  from SOHO, STEREO, SDO, and Hinode (Invited Review)}}.
\bjtitle{\solphys}
\bvolume{281},
\bfpage{187}.
\doiurl{https://doi.org/10.1007/s11207-012-9988-6}.
\adsurl{2012SoPh..281..187P}.
\end{barticle}
\endbibitem

\bibitem[\protect\citeauthoryear{{Patsourakos}, {Vourlidas}, and
  {Stenborg}}{2013}]{patsourakos2013direct}
\begin{barticle}
\bauthor{\bsnm{{Patsourakos}}, \binits{S.}},
\bauthor{\bsnm{{Vourlidas}}, \binits{A.}},
\bauthor{\bsnm{{Stenborg}}, \binits{G.}}:
\byear{2013},
\batitle{{Direct Evidence for a Fast Coronal Mass Ejection Driven by the Prior
  Formation and Subsequent Destabilization of a Magnetic Flux Rope}}.
\bjtitle{\apj}
\bvolume{764},
\bfpage{125}.
\doiurl{https://doi.org/10.1088/0004-637X/764/2/125}.
\adsurl{2013ApJ...764..125P}.
\end{barticle}
\endbibitem

\bibitem[\protect\citeauthoryear{{Priest} and
  {D{\'e}moulin}}{1995}]{priest1995three}
\begin{barticle}
\bauthor{\bsnm{{Priest}}, \binits{E.R.}},
\bauthor{\bsnm{{D{\'e}moulin}}, \binits{P.}}:
\byear{1995},
\batitle{{Three-dimensional magnetic reconnection without null points. 1. Basic
  theory of magnetic flipping}}.
\bjtitle{\jgr}
\bvolume{100},
\bfpage{23443}.
\doiurl{https://doi.org/10.1029/95JA02740}.
\adsurl{1995JGR...10023443P}.
\end{barticle}
\endbibitem

\bibitem[\protect\citeauthoryear{{Priest} and
  {Forbes}}{1992}]{priest1992magnetic}
\begin{barticle}
\bauthor{\bsnm{{Priest}}, \binits{E.R.}},
\bauthor{\bsnm{{Forbes}}, \binits{T.G.}}:
\byear{1992},
\batitle{{Magnetic Flipping: Reconnection in Three Dimensions Without Null
  Points}}.
\bjtitle{\jgr}
\bvolume{97},
\bfpage{1521}.
\doiurl{https://doi.org/10.1029/91JA02435}.
\adsurl{1992JGR....97.1521P}.
\end{barticle}
\endbibitem

\bibitem[\protect\citeauthoryear{{Priest} and
  {Longcope}}{2017}]{priest2017flux}
\begin{barticle}
\bauthor{\bsnm{{Priest}}, \binits{E.R.}},
\bauthor{\bsnm{{Longcope}}, \binits{D.W.}}:
\byear{2017},
\batitle{{Flux-Rope Twist in Eruptive Flares and CMEs: Due to Zipper and
  Main-Phase Reconnection}}.
\bjtitle{\solphys}
\bvolume{292},
\bfpage{25}.
\doiurl{https://doi.org/10.1007/s11207-016-1049-0}.
\adsurl{2017SoPh..292...25P}.
\end{barticle}
\endbibitem

\bibitem[\protect\citeauthoryear{{Savcheva}
  et~al.}{2012}]{savcheva2012photospheric}
\begin{barticle}
\bauthor{\bsnm{{Savcheva}}, \binits{A.S.}},
\bauthor{\bsnm{{Green}}, \binits{L.M.}},
\bauthor{\bsnm{{van Ballegooijen}}, \binits{A.A.}},
\bauthor{\bsnm{{DeLuca}}, \binits{E.E.}}:
\byear{2012},
\batitle{{Photospheric Flux Cancellation and the Build-up of Sigmoidal Flux
  Ropes on the Sun}}.
\bjtitle{\apj}
\bvolume{759},
\bfpage{105}.
\doiurl{https://doi.org/10.1088/0004-637X/759/2/105}.
\adsurl{2012ApJ...759..105S}.
\end{barticle}
\endbibitem

\bibitem[\protect\citeauthoryear{{Savcheva}
  et~al.}{2015}]{savcheva2015relation}
\begin{barticle}
\bauthor{\bsnm{{Savcheva}}, \binits{A.}},
\bauthor{\bsnm{{Pariat}}, \binits{E.}},
\bauthor{\bsnm{{McKillop}}, \binits{S.}},
\bauthor{\bsnm{{McCauley}}, \binits{P.}},
\bauthor{\bsnm{{Hanson}}, \binits{E.}},
\bauthor{\bsnm{{Su}}, \binits{Y.}},
\bauthor{\bsnm{{Werner}}, \binits{E.}},
\bauthor{\bsnm{{DeLuca}}, \binits{E.E.}}:
\byear{2015},
\batitle{{The Relation between Solar Eruption Topologies and Observed Flare
  Features. I. Flare Ribbons}}.
\bjtitle{\apj}
\bvolume{810},
\bfpage{96}.
\doiurl{https://doi.org/10.1088/0004-637X/810/2/96}.
\adsurl{2015ApJ...810...96S}.
\end{barticle}
\endbibitem

\bibitem[\protect\citeauthoryear{{Schmieder}, {D{\'e}moulin}, and
  {Aulanier}}{2013}]{schmieder2013solar}
\begin{barticle}
\bauthor{\bsnm{{Schmieder}}, \binits{B.}},
\bauthor{\bsnm{{D{\'e}moulin}}, \binits{P.}},
\bauthor{\bsnm{{Aulanier}}, \binits{G.}}:
\byear{2013},
\batitle{{Solar filament eruptions and their physical role in triggering
  coronal mass ejections}}.
\bjtitle{Advances in Space Research}
\bvolume{51},
\bfpage{1967}.
\doiurl{https://doi.org/10.1016/j.asr.2012.12.026}.
\adsurl{2013AdSpR..51.1967S}.
\end{barticle}
\endbibitem

\bibitem[\protect\citeauthoryear{{Scolini} et~al.}{2020}]{scolini2020cme}
\begin{barticle}
\bauthor{\bsnm{{Scolini}}, \binits{C.}},
\bauthor{\bsnm{{Chan{\'e}}}, \binits{E.}},
\bauthor{\bsnm{{Temmer}}, \binits{M.}},
\bauthor{\bsnm{{Kilpua}}, \binits{E.K.J.}},
\bauthor{\bsnm{{Dissauer}}, \binits{K.}},
\bauthor{\bsnm{{Veronig}}, \binits{A.M.}},
\bauthor{\bsnm{{Palmerio}}, \binits{E.}},
\bauthor{\bsnm{{Pomoell}}, \binits{J.}},
\bauthor{\bsnm{{Dumbovi{\'c}}}, \binits{M.}},
\bauthor{\bsnm{{Guo}}, \binits{J.}},
\bauthor{\bsnm{{Rodriguez}}, \binits{L.}},
\bauthor{\bsnm{{Poedts}}, \binits{S.}}:
\byear{2020},
\batitle{{CME-CME Interactions as Sources of CME Geoeffectiveness: The
  Formation of the Complex Ejecta and Intense Geomagnetic Storm in 2017 Early
  September}}.
\bjtitle{\apj\ Suppl.}
\bvolume{247},
\bfpage{21}.
\doiurl{https://doi.org/10.3847/1538-4365/ab6216}.
\adsurl{2020ApJS..247...21S}.
\end{barticle}
\endbibitem

\bibitem[\protect\citeauthoryear{{SONG} et~al.}{2015}]{song2015evidence}
\begin{barticle}
\bauthor{\bsnm{{SONG}}, \binits{H.Q.}},
\bauthor{\bsnm{{CHEN}}, \binits{Y.}},
\bauthor{\bsnm{{ZHANG}}, \binits{J.}},
\bauthor{\bsnm{{CHENG}}, \binits{X.}},
\bauthor{\bsnm{{Wang}}, \binits{B.}},
\bauthor{\bsnm{{HU}}, \binits{Q.}},
\bauthor{\bsnm{{LI}}, \binits{G.}},
\bauthor{\bsnm{{WANG}}, \binits{Y.M.}}:
\byear{2015},
\batitle{{Evidence of the Solar EUV Hot Channel as a Magnetic Flux Rope from
  Remote-sensing and In Situ Observations}}.
\bjtitle{\apjl}
\bvolume{808},
\bfpage{L15}.
\doiurl{https://doi.org/10.1088/2041-8205/808/1/L15}.
\adsurl{2015ApJ...808L..15S}.
\end{barticle}
\endbibitem

\bibitem[\protect\citeauthoryear{{Sun} et~al.}{2022}]{sun2022cross}
\begin{barticle}
\bauthor{\bsnm{{Sun}}, \binits{Z.}},
\bauthor{\bsnm{{Tian}}, \binits{H.}},
\bauthor{\bsnm{{Chen}}, \binits{P.F.}},
\bauthor{\bsnm{{Yao}}, \binits{S.}},
\bauthor{\bsnm{{Hou}}, \binits{Z.}},
\bauthor{\bsnm{{Chen}}, \binits{H.}},
\bauthor{\bsnm{{Chen}}, \binits{L.}}:
\byear{2022},
\batitle{{Cross-loop Propagation of a Quasiperiodic Extreme-ultraviolet Wave
  Train Triggered by Successive Stretching of Magnetic Field Structures during
  a Solar Eruption}}.
\bjtitle{\apjl}
\bvolume{939},
\bfpage{L18}.
\doiurl{https://doi.org/10.3847/2041-8213/ac9aff}.
\adsurl{2022ApJ...939L..18S}.
\end{barticle}
\endbibitem

\bibitem[\protect\citeauthoryear{{Sun} et~al.}{2023}]{sun2023observation}
\begin{barticle}
\bauthor{\bsnm{{Sun}}, \binits{Z.}},
\bauthor{\bsnm{{Li}}, \binits{T.}},
\bauthor{\bsnm{{Tian}}, \binits{H.}},
\bauthor{\bsnm{{Hou}}, \binits{Y.}},
\bauthor{\bsnm{{Hou}}, \binits{Z.}},
\bauthor{\bsnm{{Chen}}, \binits{H.}},
\bauthor{\bsnm{{Bai}}, \binits{X.}},
\bauthor{\bsnm{{Deng}}, \binits{Y.}}:
\byear{2023},
\batitle{{Observation of Two Splitting Processes in a Partial Filament Eruption
  on the Sun: The Role of Breakout Reconnection}}.
\bjtitle{\apj}
\bvolume{953},
\bfpage{148}.
\doiurl{https://doi.org/10.3847/1538-4357/ace5b1}.
\adsurl{2023ApJ...953..148S}.
\end{barticle}
\endbibitem

\bibitem[\protect\citeauthoryear{{Tian}}{2017}]{tian2017probing}
\begin{barticle}
\bauthor{\bsnm{{Tian}}, \binits{H.}}:
\byear{2017},
\batitle{{Probing the solar transition region: current status and future
  perspectives}}.
\bjtitle{Research in Astronomy and Astrophysics}
\bvolume{17},
\bfpage{110}.
\doiurl{https://doi.org/10.1088/1674-4527/17/11/110}.
\adsurl{2017RAA....17..110T}.
\end{barticle}
\endbibitem

\bibitem[\protect\citeauthoryear{{Titov} and
  {D{\'e}moulin}}{1999}]{titov1999basic}
\begin{barticle}
\bauthor{\bsnm{{Titov}}, \binits{V.S.}},
\bauthor{\bsnm{{D{\'e}moulin}}, \binits{P.}}:
\byear{1999},
\batitle{{Basic topology of twisted magnetic configurations in solar flares}}.
\bjtitle{\aap}
\bvolume{351},
\bfpage{707}.
\adsurl{1999A&A...351..707T}.
\end{barticle}
\endbibitem

\bibitem[\protect\citeauthoryear{{Titov}, {Hornig}, and
  {D{\'e}moulin}}{2002}]{titov2002theory}
\begin{barticle}
\bauthor{\bsnm{{Titov}}, \binits{V.S.}},
\bauthor{\bsnm{{Hornig}}, \binits{G.}},
\bauthor{\bsnm{{D{\'e}moulin}}, \binits{P.}}:
\byear{2002},
\batitle{{Theory of magnetic connectivity in the solar corona}}.
\bjtitle{Journal of Geophysical Research (Space Physics)}
\bvolume{107},
\bfpage{1164}.
\doiurl{https://doi.org/10.1029/2001JA000278}.
\adsurl{2002JGRA..107.1164T}.
\end{barticle}
\endbibitem

\bibitem[\protect\citeauthoryear{{Toriumi} and {Wang}}{2019}]{toriumi2019flare}
\begin{barticle}
\bauthor{\bsnm{{Toriumi}}, \binits{S.}},
\bauthor{\bsnm{{Wang}}, \binits{H.}}:
\byear{2019},
\batitle{{Flare-productive active regions}}.
\bjtitle{Living Reviews in Solar Physics}
\bvolume{16},
\bfpage{3}.
\doiurl{https://doi.org/10.1007/s41116-019-0019-7}.
\adsurl{2019LRSP...16....3T}.
\end{barticle}
\endbibitem

\bibitem[\protect\citeauthoryear{{T{\"o}r{\"o}k}
  et~al.}{2014}]{torok2014distribution}
\begin{barticle}
\bauthor{\bsnm{{T{\"o}r{\"o}k}}, \binits{T.}},
\bauthor{\bsnm{{Leake}}, \binits{J.E.}},
\bauthor{\bsnm{{Titov}}, \binits{V.S.}},
\bauthor{\bsnm{{Archontis}}, \binits{V.}},
\bauthor{\bsnm{{Miki{\'c}}}, \binits{Z.}},
\bauthor{\bsnm{{Linton}}, \binits{M.G.}},
\bauthor{\bsnm{{Dalmasse}}, \binits{K.}},
\bauthor{\bsnm{{Aulanier}}, \binits{G.}},
\bauthor{\bsnm{{Kliem}}, \binits{B.}}:
\byear{2014},
\batitle{{Distribution of Electric Currents in Solar Active Regions}}.
\bjtitle{\apjl}
\bvolume{782},
\bfpage{L10}.
\doiurl{https://doi.org/10.1088/2041-8205/782/1/L10}.
\adsurl{2014ApJ...782L..10T}.
\end{barticle}
\endbibitem

\bibitem[\protect\citeauthoryear{{Veronig} et~al.}{2018}]{veronig2018genesis}
\begin{barticle}
\bauthor{\bsnm{{Veronig}}, \binits{A.M.}},
\bauthor{\bsnm{{Podladchikova}}, \binits{T.}},
\bauthor{\bsnm{{Dissauer}}, \binits{K.}},
\bauthor{\bsnm{{Temmer}}, \binits{M.}},
\bauthor{\bsnm{{Seaton}}, \binits{D.B.}},
\bauthor{\bsnm{{Long}}, \binits{D.}},
\bauthor{\bsnm{{Guo}}, \binits{J.}},
\bauthor{\bsnm{{Vr{\v{s}}nak}}, \binits{B.}},
\bauthor{\bsnm{{Harra}}, \binits{L.}},
\bauthor{\bsnm{{Kliem}}, \binits{B.}}:
\byear{2018},
\batitle{{Genesis and Impulsive Evolution of the 2017 September 10 Coronal Mass
  Ejection}}.
\bjtitle{\apj}
\bvolume{868},
\bfpage{107}.
\doiurl{https://doi.org/10.3847/1538-4357/aaeac5}.
\adsurl{2018ApJ...868..107V}.
\end{barticle}
\endbibitem

\bibitem[\protect\citeauthoryear{{Wang} et~al.}{2017}]{wang2017buildup}
\begin{barticle}
\bauthor{\bsnm{{Wang}}, \binits{W.}},
\bauthor{\bsnm{{Liu}}, \binits{R.}},
\bauthor{\bsnm{{Wang}}, \binits{Y.}},
\bauthor{\bsnm{{Hu}}, \binits{Q.}},
\bauthor{\bsnm{{Shen}}, \binits{C.}},
\bauthor{\bsnm{{Jiang}}, \binits{C.}},
\bauthor{\bsnm{{Zhu}}, \binits{C.}}:
\byear{2017},
\batitle{{Buildup of a highly twisted magnetic flux rope during a solar
  eruption}}.
\bjtitle{Nature Communications}
\bvolume{8},
\bfpage{1330}.
\doiurl{https://doi.org/10.1038/s41467-017-01207-x}.
\adsurl{2017NatCo...8.1330W}.
\end{barticle}
\endbibitem

\bibitem[\protect\citeauthoryear{{Warmuth}}{2015}]{warmuth2015large}
\begin{barticle}
\bauthor{\bsnm{{Warmuth}}, \binits{A.}}:
\byear{2015},
\batitle{{Large-scale Globally Propagating Coronal Waves}}.
\bjtitle{Living Reviews in Solar Physics}
\bvolume{12},
\bfpage{3}.
\doiurl{https://doi.org/10.1007/lrsp-2015-3}.
\adsurl{2015LRSP...12....3W}.
\end{barticle}
\endbibitem

\bibitem[\protect\citeauthoryear{{Wheatland}, {Sturrock}, and
  {Roumeliotis}}{2000}]{wheatland2000optimization}
\begin{barticle}
\bauthor{\bsnm{{Wheatland}}, \binits{M.S.}},
\bauthor{\bsnm{{Sturrock}}, \binits{P.A.}},
\bauthor{\bsnm{{Roumeliotis}}, \binits{G.}}:
\byear{2000},
\batitle{{An Optimization Approach to Reconstructing Force-free Fields}}.
\bjtitle{\apj}
\bvolume{540},
\bfpage{1150}.
\doiurl{https://doi.org/10.1086/309355}.
\adsurl{2000ApJ...540.1150W}.
\end{barticle}
\endbibitem

\bibitem[\protect\citeauthoryear{{Wiegelmann}}{2004}]{wiegelmann2004optimization}
\begin{barticle}
\bauthor{\bsnm{{Wiegelmann}}, \binits{T.}}:
\byear{2004},
\batitle{{Optimization code with weighting function for the reconstruction of
  coronal magnetic fields}}.
\bjtitle{\solphys}
\bvolume{219},
\bfpage{87}.
\doiurl{https://doi.org/10.1023/B:SOLA.0000021799.39465.36}.
\adsurl{2004SoPh..219...87W}.
\end{barticle}
\endbibitem

\bibitem[\protect\citeauthoryear{{Wiegelmann}, {Inhester}, and
  {Sakurai}}{2006}]{wiegelmann2006preprocessing}
\begin{barticle}
\bauthor{\bsnm{{Wiegelmann}}, \binits{T.}},
\bauthor{\bsnm{{Inhester}}, \binits{B.}},
\bauthor{\bsnm{{Sakurai}}, \binits{T.}}:
\byear{2006},
\batitle{{Preprocessing of Vector Magnetograph Data for a Nonlinear Force-Free
  Magnetic Field Reconstruction}}.
\bjtitle{\solphys}
\bvolume{233},
\bfpage{215}.
\doiurl{https://doi.org/10.1007/s11207-006-2092-z}.
\adsurl{2006SoPh..233..215W}.
\end{barticle}
\endbibitem

\bibitem[\protect\citeauthoryear{{Wiegelmann}
  et~al.}{2012}]{wiegelmann2012should}
\begin{barticle}
\bauthor{\bsnm{{Wiegelmann}}, \binits{T.}},
\bauthor{\bsnm{{Thalmann}}, \binits{J.K.}},
\bauthor{\bsnm{{Inhester}}, \binits{B.}},
\bauthor{\bsnm{{Tadesse}}, \binits{T.}},
\bauthor{\bsnm{{Sun}}, \binits{X.}},
\bauthor{\bsnm{{Hoeksema}}, \binits{J.T.}}:
\byear{2012},
\batitle{{How Should One Optimize Nonlinear Force-Free Coronal Magnetic Field
  Extrapolations from SDO/HMI Vector Magnetograms?}}
\bjtitle{\solphys}
\bvolume{281},
\bfpage{37}.
\doiurl{https://doi.org/10.1007/s11207-012-9966-z}.
\adsurl{2012SoPh..281...37W}.
\end{barticle}
\endbibitem

\bibitem[\protect\citeauthoryear{{Xie} et~al.}{2023}]{xie2023magnetic}
\begin{barticle}
\bauthor{\bsnm{{Xie}}, \binits{H.}},
\bauthor{\bsnm{{Gopalswamy}}, \binits{N.}},
\bauthor{\bsnm{{Akiyama}}, \binits{S.}},
\bauthor{\bsnm{{Yashiro}}, \binits{S.}},
\bauthor{\bsnm{{Makela}}, \binits{P.}}:
\byear{2023},
\batitle{{Magnetic flux rope structures associated with filament channels: Two
  case studies}}.
\bjtitle{Journal of Atmospheric and Solar-Terrestrial Physics}
\bvolume{252},
\bfpage{106154}.
\doiurl{https://doi.org/10.1016/j.jastp.2023.106154}.
\adsurl{2023JASTP.25206154X}.
\end{barticle}
\endbibitem

\bibitem[\protect\citeauthoryear{{Zhang}, {Cheng}, and
  {Ding}}{2012}]{zhang2012observation}
\begin{barticle}
\bauthor{\bsnm{{Zhang}}, \binits{J.}},
\bauthor{\bsnm{{Cheng}}, \binits{X.}},
\bauthor{\bsnm{{Ding}}, \binits{M.-D.}}:
\byear{2012},
\batitle{{Observation of an evolving magnetic flux rope before and during a
  solar eruption}}.
\bjtitle{Nature Communications}
\bvolume{3},
\bfpage{747}.
\doiurl{https://doi.org/10.1038/ncomms1753}.
\adsurl{2012NatCo...3..747Z}.
\end{barticle}
\endbibitem

\bibitem[\protect\citeauthoryear{{Zhang} et~al.}{2007}]{zhang2007solar}
\begin{barticle}
\bauthor{\bsnm{{Zhang}}, \binits{J.}},
\bauthor{\bsnm{{Richardson}}, \binits{I.G.}},
\bauthor{\bsnm{{Webb}}, \binits{D.F.}},
\bauthor{\bsnm{{Gopalswamy}}, \binits{N.}},
\bauthor{\bsnm{{Huttunen}}, \binits{E.}},
\bauthor{\bsnm{{Kasper}}, \binits{J.C.}},
\bauthor{\bsnm{{Nitta}}, \binits{N.V.}},
\bauthor{\bsnm{{Poomvises}}, \binits{W.}},
\bauthor{\bsnm{{Thompson}}, \binits{B.J.}},
\bauthor{\bsnm{{Wu}}, \binits{C.-C.}},
\bauthor{\bsnm{{Yashiro}}, \binits{S.}},
\bauthor{\bsnm{{Zhukov}}, \binits{A.N.}}:
\byear{2007},
\batitle{{Solar and interplanetary sources of major geomagnetic storms (Dst <=
  -100 nT) during 1996-2005}}.
\bjtitle{Journal of Geophysical Research (Space Physics)}
\bvolume{112},
\bfpage{A10102}.
\doiurl{https://doi.org/10.1029/2007JA012321}.
\adsurl{2007JGRA..11210102Z}.
\end{barticle}
\endbibitem

\bibitem[\protect\citeauthoryear{{Zhang} et~al.}{2019}]{zhang2019hard}
\begin{barticle}
\bauthor{\bsnm{{Zhang}}, \binits{Z.}},
\bauthor{\bsnm{{Chen}}, \binits{D.-Y.}},
\bauthor{\bsnm{{Wu}}, \binits{J.}},
\bauthor{\bsnm{{Chang}}, \binits{J.}},
\bauthor{\bsnm{{Hu}}, \binits{Y.-M.}},
\bauthor{\bsnm{{Su}}, \binits{Y.}},
\bauthor{\bsnm{{Zhang}}, \binits{Y.}},
\bauthor{\bsnm{{Wang}}, \binits{J.-P.}},
\bauthor{\bsnm{{Liang}}, \binits{Y.-M.}},
\bauthor{\bsnm{{Ma}}, \binits{T.}},
\bauthor{\bsnm{{Guo}}, \binits{J.-H.}},
\bauthor{\bsnm{{Cai}}, \binits{M.-S.}},
\bauthor{\bsnm{{Zhang}}, \binits{Y.-Q.}},
\bauthor{\bsnm{{Huang}}, \binits{Y.-Y.}},
\bauthor{\bsnm{{Peng}}, \binits{X.-Y.}},
\bauthor{\bsnm{{Tang}}, \binits{Z.-B.}},
\bauthor{\bsnm{{Zhao}}, \binits{X.}},
\bauthor{\bsnm{{Zhou}}, \binits{H.-H.}},
\bauthor{\bsnm{{Wang}}, \binits{L.-G.}},
\bauthor{\bsnm{{Song}}, \binits{J.-X.}},
\bauthor{\bsnm{{Ma}}, \binits{M.}},
\bauthor{\bsnm{{Xu}}, \binits{G.-Z.}},
\bauthor{\bsnm{{Yang}}, \binits{J.-F.}},
\bauthor{\bsnm{{Lu}}, \binits{D.}},
\bauthor{\bsnm{{He}}, \binits{Y.-H.}},
\bauthor{\bsnm{{Tao}}, \binits{J.-Y.}},
\bauthor{\bsnm{{Ma}}, \binits{X.-L.}},
\bauthor{\bsnm{{Lv}}, \binits{B.-G.}},
\bauthor{\bsnm{{Bai}}, \binits{Y.-P.}},
\bauthor{\bsnm{{Cao}}, \binits{C.-X.}},
\bauthor{\bsnm{{Huang}}, \binits{Y.}},
\bauthor{\bsnm{{Gan}}, \binits{W.-Q.}}:
\byear{2019},
\batitle{{Hard X-ray Imager (HXI) onboard the ASO-S mission}}.
\bjtitle{Research in Astronomy and Astrophysics}
\bvolume{19},
\bfpage{160}.
\doiurl{https://doi.org/10.1088/1674-4527/19/11/160}.
\adsurl{2019RAA....19..160Z}.
\end{barticle}
\endbibitem

\bibitem[\protect\citeauthoryear{{Zhao} et~al.}{2016}]{zhao2016hooked}
\begin{barticle}
\bauthor{\bsnm{{Zhao}}, \binits{J.}},
\bauthor{\bsnm{{Gilchrist}}, \binits{S.A.}},
\bauthor{\bsnm{{Aulanier}}, \binits{G.}},
\bauthor{\bsnm{{Schmieder}}, \binits{B.}},
\bauthor{\bsnm{{Pariat}}, \binits{E.}},
\bauthor{\bsnm{{Li}}, \binits{H.}}:
\byear{2016},
\batitle{{Hooked Flare Ribbons and Flux-rope-related QSL Footprints}}.
\bjtitle{\apj}
\bvolume{823},
\bfpage{62}.
\doiurl{https://doi.org/10.3847/0004-637X/823/1/62}.
\adsurl{2016ApJ...823...62Z}.
\end{barticle}
\endbibitem

\bibitem[\protect\citeauthoryear{{Zheng} et~al.}{2021}]{zheng2021compound}
\begin{barticle}
\bauthor{\bsnm{{Zheng}}, \binits{R.}},
\bauthor{\bsnm{{Zhang}}, \binits{L.}},
\bauthor{\bsnm{{Wang}}, \binits{B.}},
\bauthor{\bsnm{{Kong}}, \binits{X.}},
\bauthor{\bsnm{{Song}}, \binits{H.}},
\bauthor{\bsnm{{Wu}}, \binits{Z.}},
\bauthor{\bsnm{{Feng}}, \binits{S.}},
\bauthor{\bsnm{{Chen}}, \binits{H.}},
\bauthor{\bsnm{{Chen}}, \binits{Y.}}:
\byear{2021},
\batitle{{Compound Eruptions of Twin Flux Ropes in a Solar Active Region}}.
\bjtitle{\apjl}
\bvolume{921},
\bfpage{L39}.
\doiurl{https://doi.org/10.3847/2041-8213/ac33ae}.
\adsurl{2021ApJ...921L..39Z}.
\end{barticle}
\endbibitem

\bibitem[\protect\citeauthoryear{{Zhongxian} and
  {Jingxiu}}{1994}]{shi1994delta}
\begin{barticle}
\bauthor{\bsnm{{Zhongxian}}, \binits{S.}},
\bauthor{\bsnm{{Jingxiu}}, \binits{W.}}:
\byear{1994},
\batitle{{Delta-Sunspots and X-Class Flares}}.
\bjtitle{\solphys}
\bvolume{149},
\bfpage{105}.
\doiurl{https://doi.org/10.1007/BF00645181}.
\adsurl{1994SoPh..149..105S}.
\end{barticle}
\endbibitem

\end{thebibliography}

\newpage
\begin{figure}
  \centering
  \includegraphics[width=1.0\textwidth]{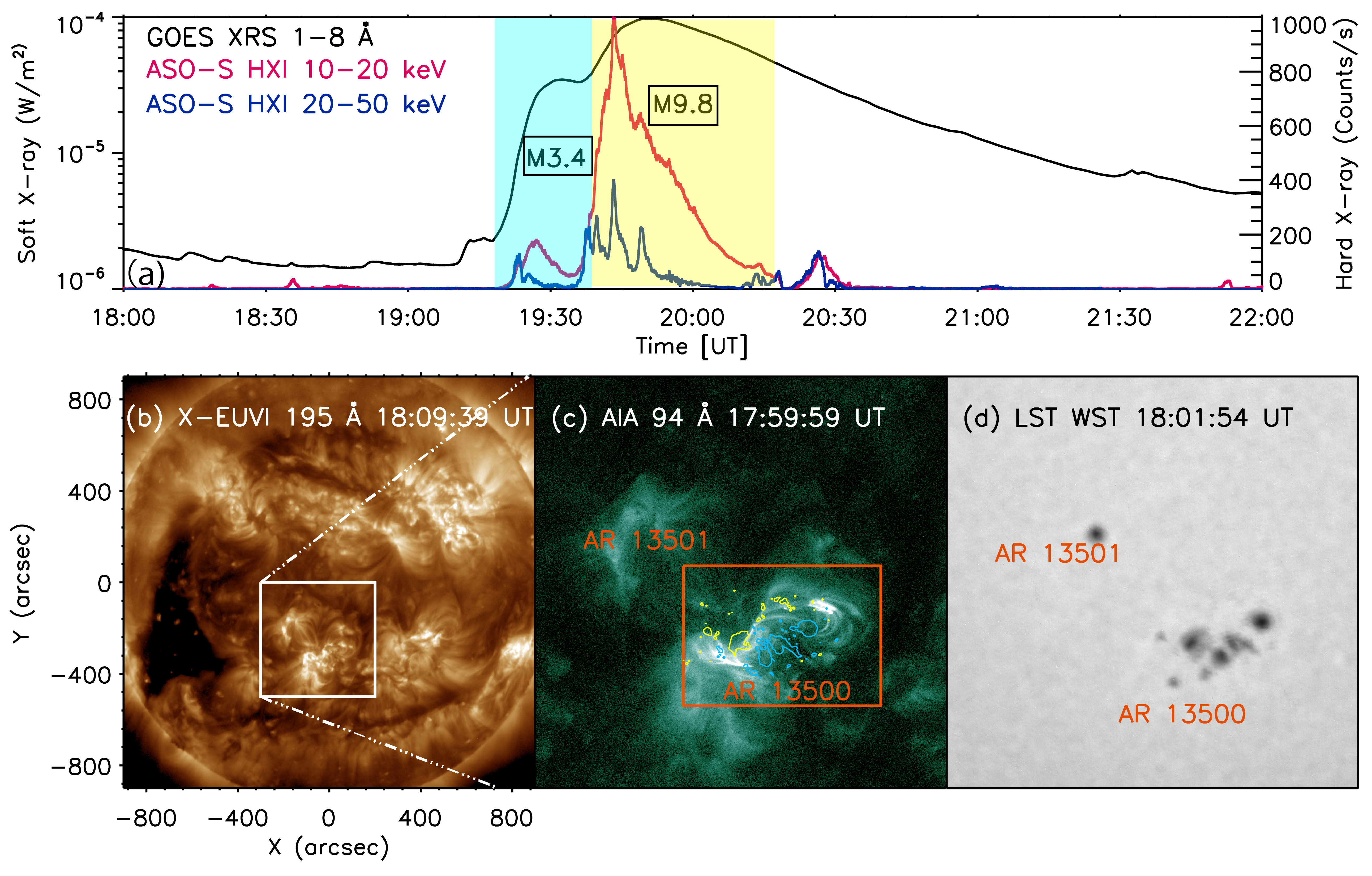}
\caption{Overview of the eruption. panel (a): temporal evolution of the GOES XRS 1-8~\AA\ (black), ASO-S HXI 10-20 keV (red), and 20-50 keV (blue). The blue and yellow rectangles indicate the time range of the M3.4-class flare and the M9.8-class flare, respectively. panel (b): overview of the AR 13501 and AR 13500 in FY-3E/X-EUVI 195~\AA. The white rectangle denotes the field of view (FOV) of panels (c)-(d). Panels (c)-(d): zoom-in area of AR 13500 and AR 13501 in AIA 94~\AA\ (c) and in LST/ WST 3600~\AA\ (d). The orange rectangle shows the FOV of Figure 2. Panel (c) is superimposed by contours of the magnetic field at $\pm$ 400 Gauss levels, with yellow indicating positive values and blue representing negative values.
  \label{fig:Fig.1}}
\end{figure}

\begin{figure}
  \centering
  \includegraphics[width=1.0\textwidth]{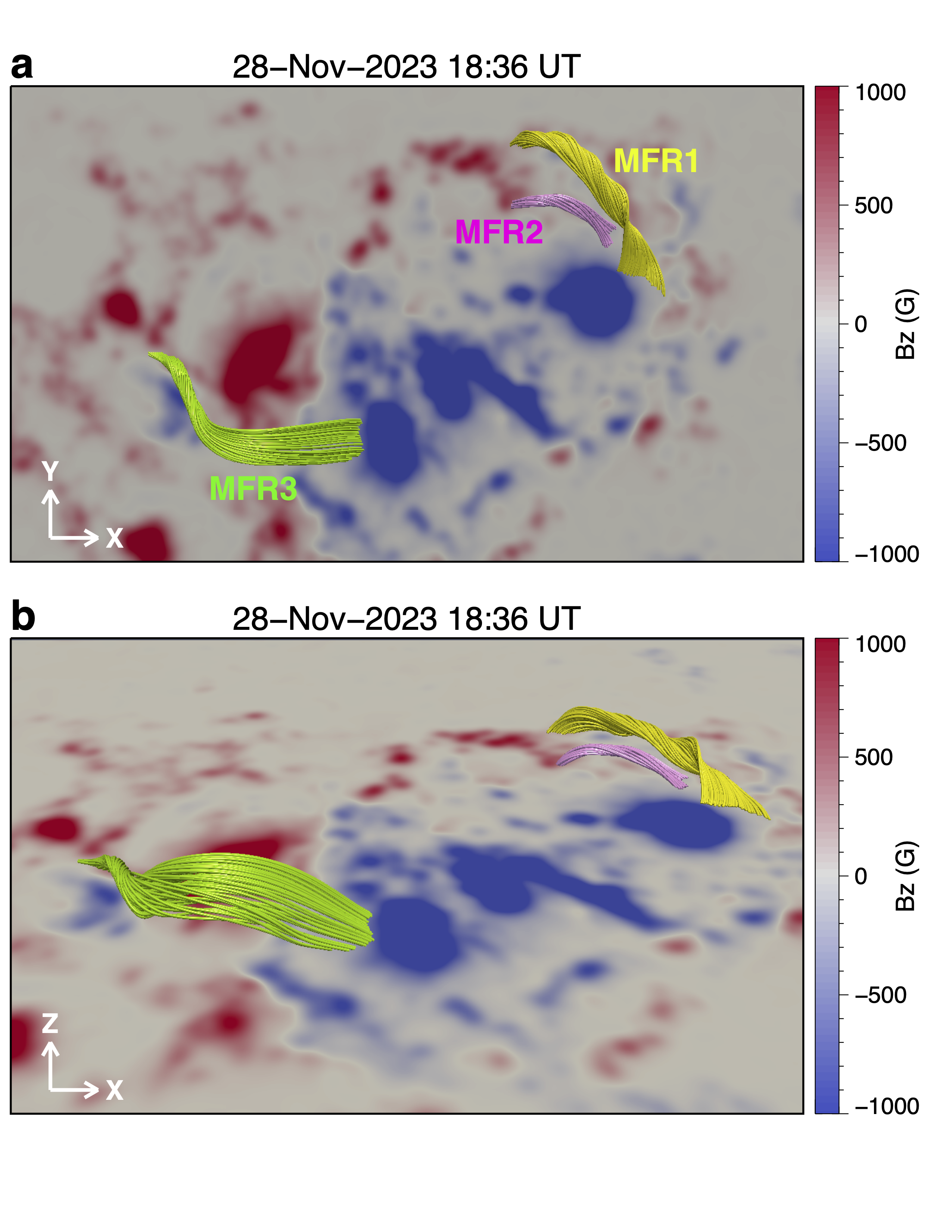}
  \caption{Multi-MFR configuration of the AR 13500 revealed by NLFFF extrapolation at 18:36 UT. Panel (a) and (b) shows the top view and side view of the extrapolated results, respectively. Three identified MFRs are marked as MFR1, MFR2 and MFR3. The extrapolation is performed within the area with a twist number above 1.0.}
  \label{fig:Fig.2}
\end{figure}

\begin{figure}
  \centering
  \includegraphics[width=1.0\textwidth]{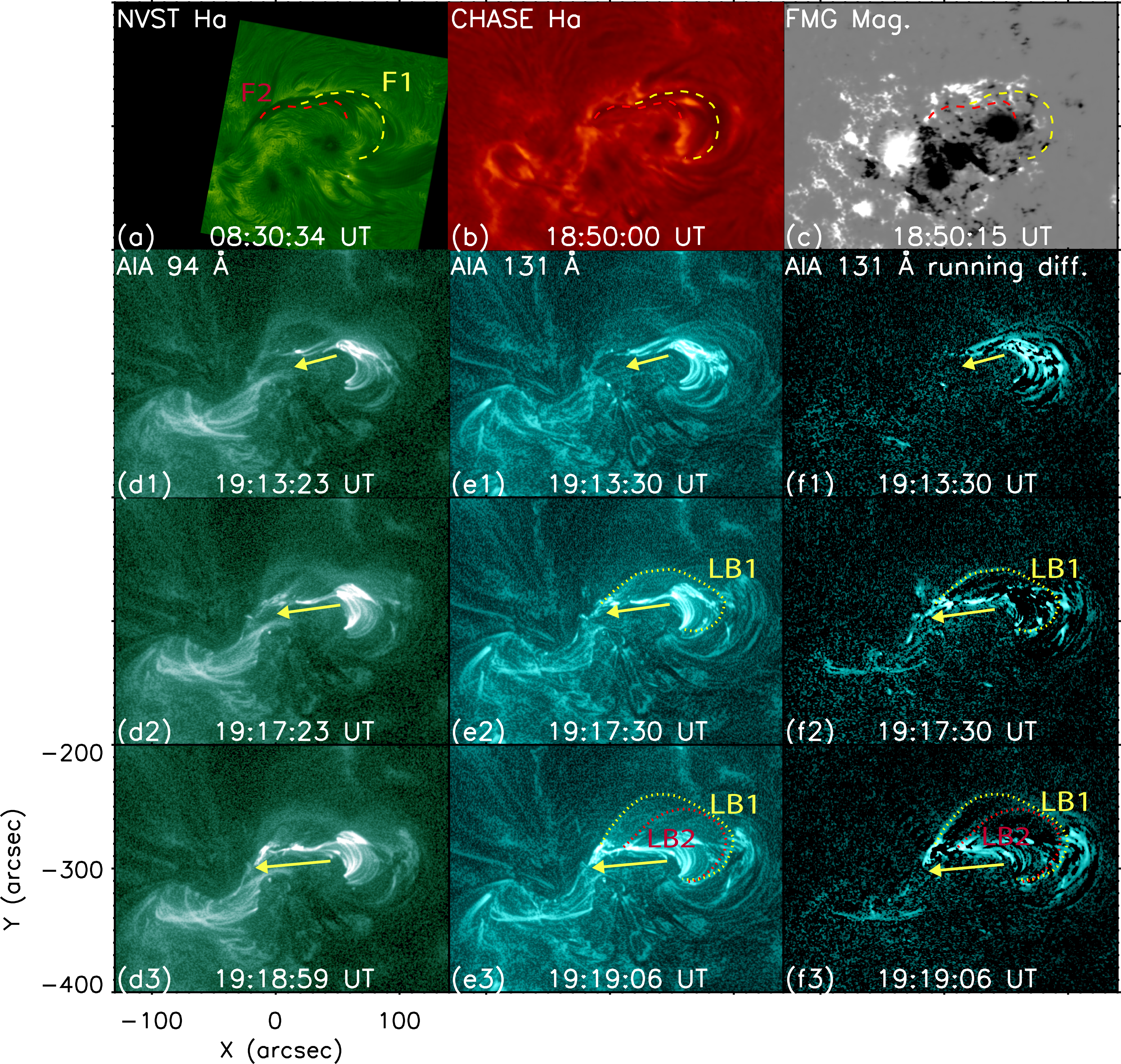}
  \caption{Pre-eruption observations and slipping process of MFR1. Panels (a) and (b) shows the H$\alpha$ observations from NVST and CHASE, respectively. Panels (c) displays the FMG magnetogram. The red and yellow dashes lines depict the location of Filament 1 (F1) and Filament 2 (F2), respectively. Panels (d)-(f) shows the EUV observations of the slipping process of MFR1 in 94~\AA\ (panel (d)), 131~\AA\ (panel (e)), and 131~\AA\ running difference with a 48-second cadence. The yellow arrow shows the slipping process of MFR1, while the yellow and red dotted lines represent the loop bundles of LB1 and LB2, respectively. The LB1 and LB2 might be associated with MFR1 and MFR2, respectively. (An animation of this figure is available online.)
  \label{fig:Fig.3}}
\end{figure}

\begin{figure}
  \centering
  \includegraphics[width=1.0\textwidth]{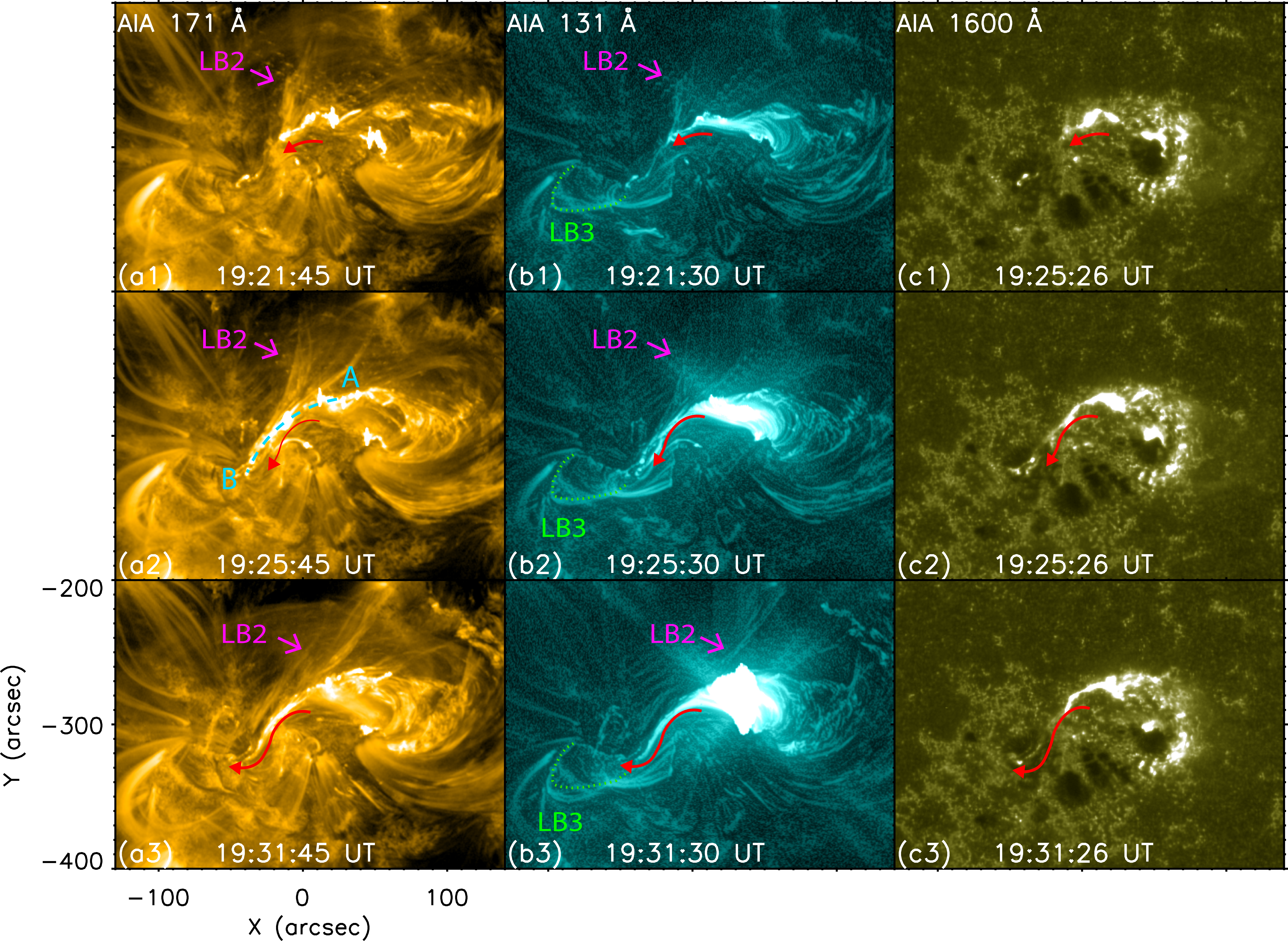}
  \caption{Slipping process of MFR2. Panels (a)-(c) display the observations in AIA 171~\AA, 131~\AA, and 1600~\AA, respectively. The pink arrow denotes the location of the outflows resulting from the reconnection beneath the MFR, thereby aiding in the identification of MFR2. The red arrow shows the slipping process of MFR2. Additionally, the green dots indicate the loop bundle LB3, which might be associated with MFR3. Slice A-B in panel (a2) is specifically selected to illustrate the slipping process of MFR1 and MFR2 and the result is shown in Figure 6(a). The observed solar flare is the M3.4-class flare produced by the eruption of MFR1.
  \label{fig:Fig.4}}
\end{figure}

\begin{figure}
  \centering
  \includegraphics[width=1.0\textwidth]{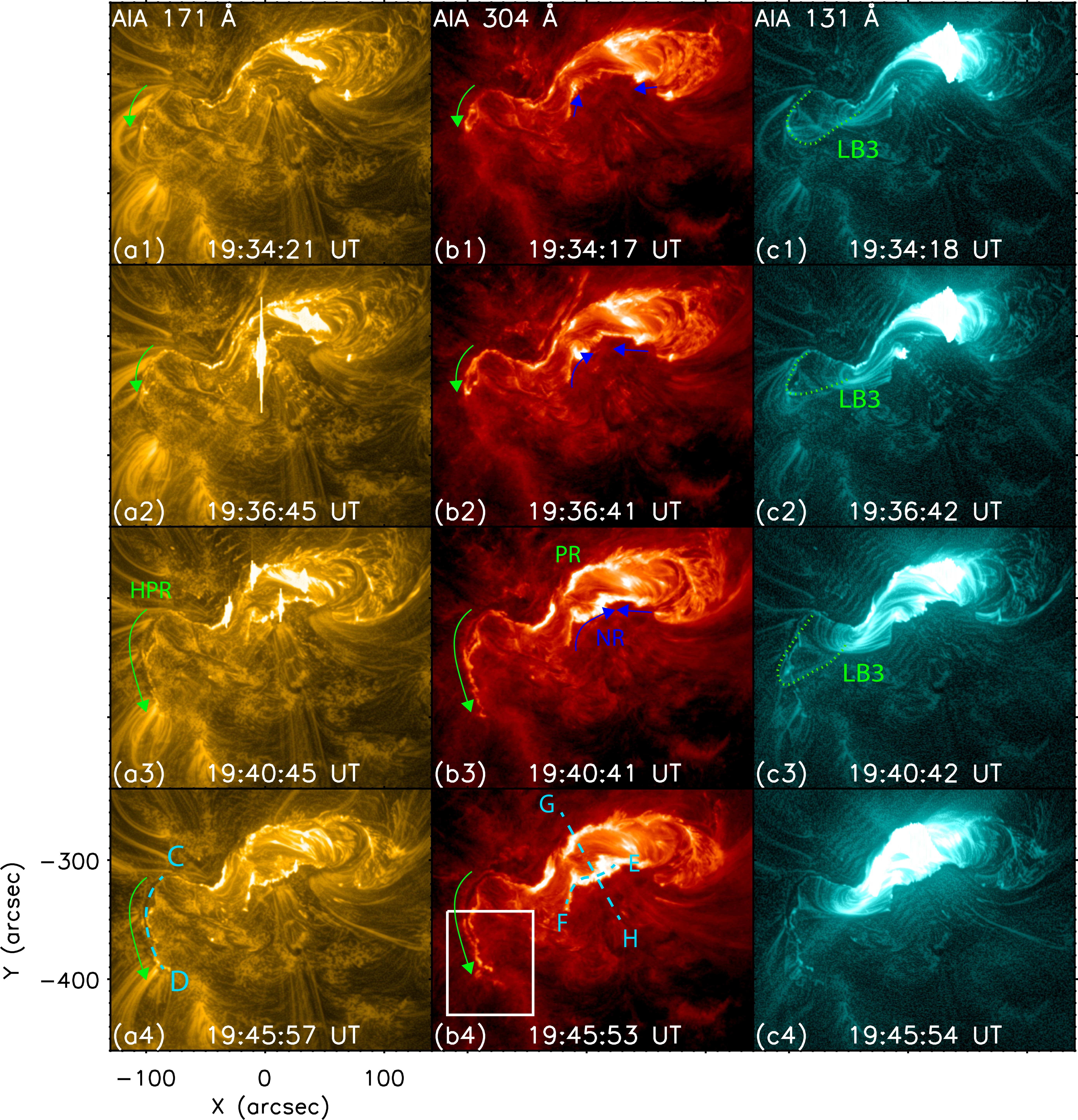}
  \caption{Slipping process of MFR3. Panels (a)-(c) depict the observations in AIA 171~\AA, 304~\AA, and 131~\AA, respectively. The green arrow illustrates the slipping process of LB3, while the blue arrows indicate the bi-directional slipping of the second flare ribbon. Green dots in panels (c1)-(c3) mark the location of MFR3's loop bundle. The white rectangle in panel (b4) outlines the FOV depicted in Figure 7. Slices C-D, E-F, and G-H in the bottom row are specifically chosen to elucidate the slipping process of MFR3, bi-directional slipping of the second ribbon, and the separation of the double flare ribbons, respectively. The observed flare corresponds to the M9.8-class flare resulting from the eruption of MFR2 and MFR3.
  \label{fig:Fig.5}}
\end{figure}

\begin{figure}
  \centering
  \includegraphics[width=1.0\textwidth]{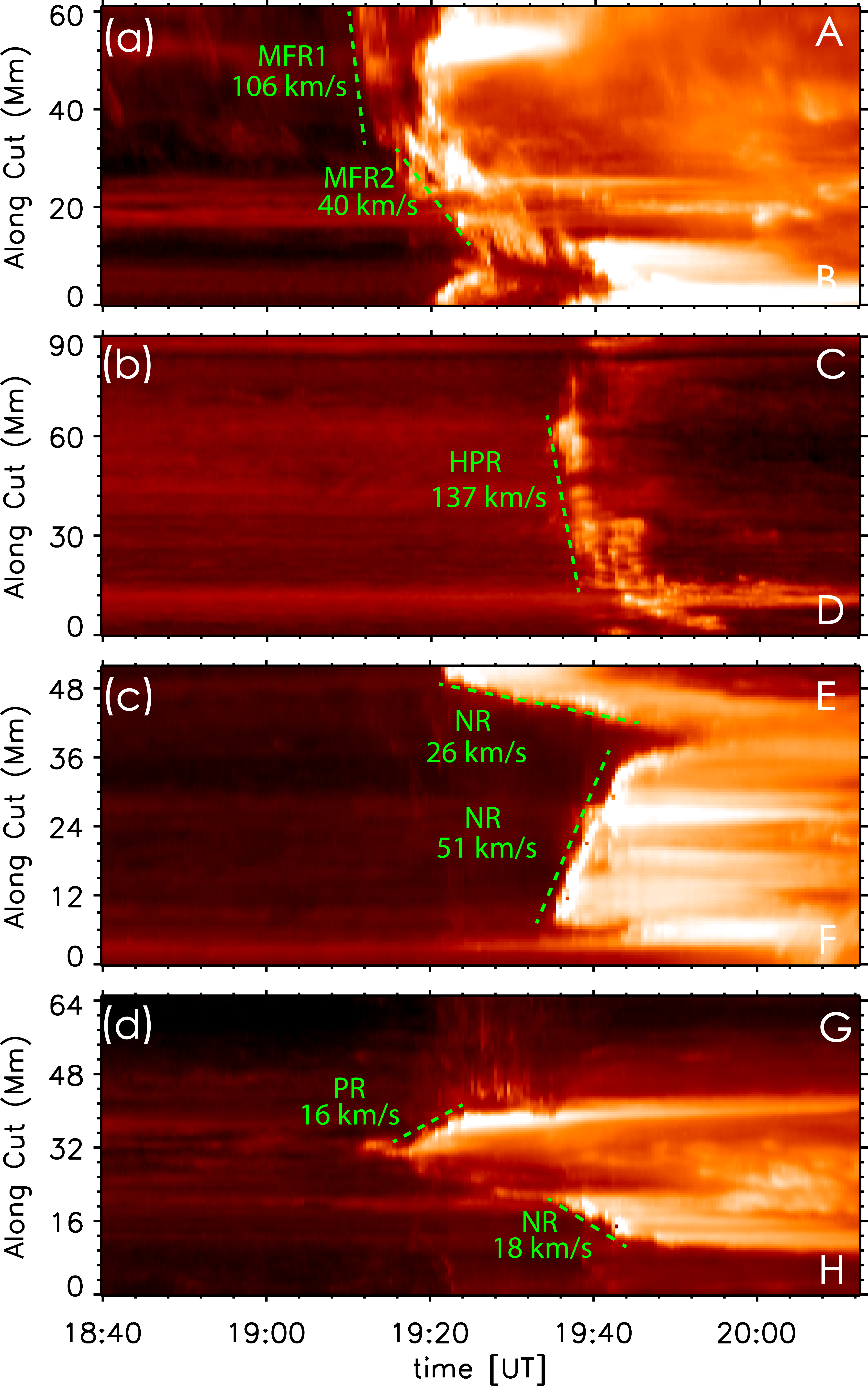}
  \caption{Time-distance diagrams. Slice A-B depicts the slipping process of MFR1 and MFR2, while slice C-D illustrates the slipping process of MFR3. Slice E-F shows the bi-directional slipping of the second ribbon, and slice G-H illustrates the separation of the double flare ribbons. Slice A-B and C-D are shown in the Figure 4(a2) and Figure 5(a4), respectively, while Slice E-F and G-H are displayed in Figure 5(b4). The calculated velocity is denoted in green and all velocities are projected velocities.}
  \label{fig:Fig.6}
\end{figure}

\begin{figure}
  \centering
  \includegraphics[width=1.0\textwidth]{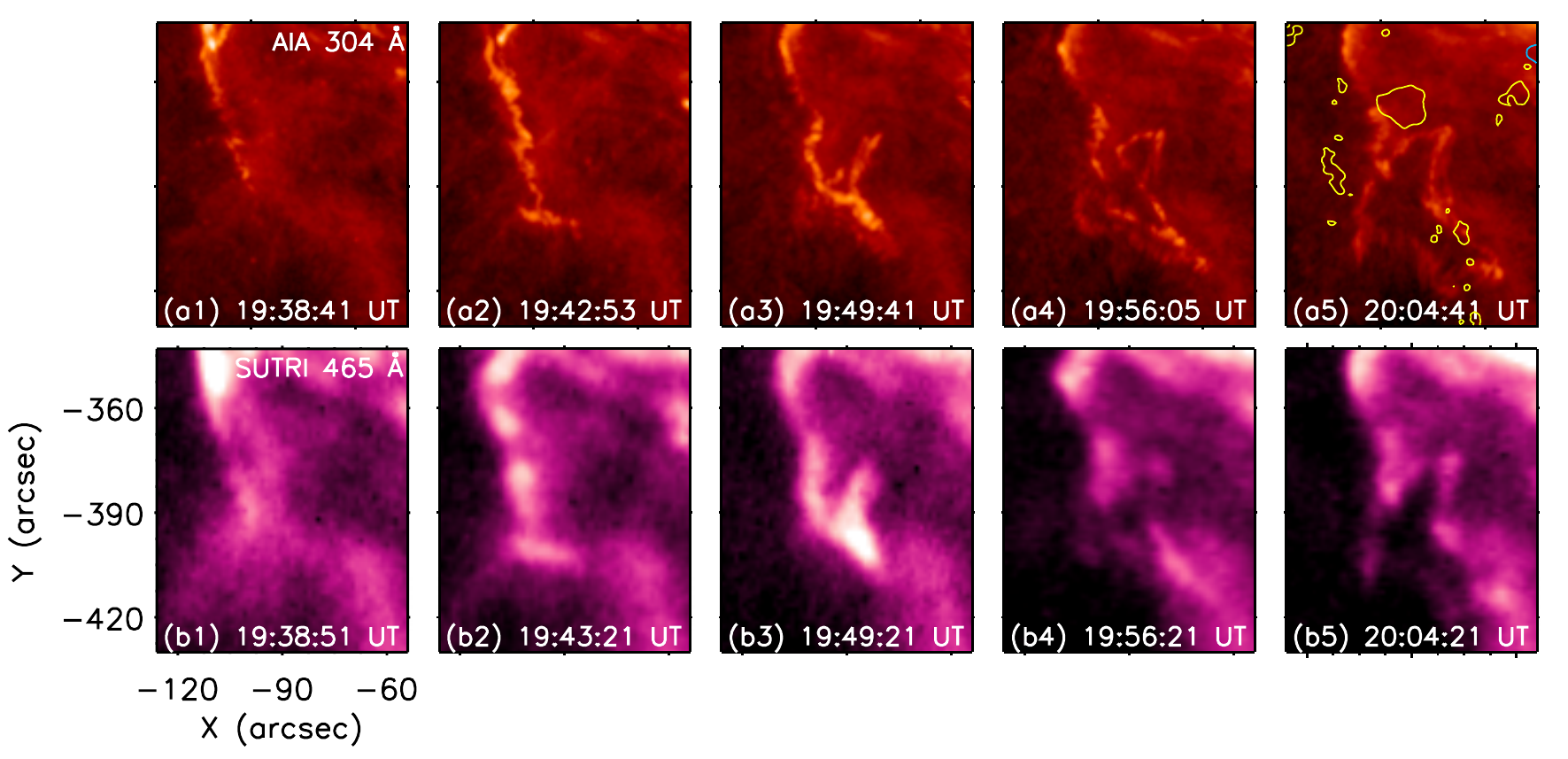}
  \caption{Fine structures of the end of the flare ribbon in the slipping process of MFR3, observed in AIA 304~\AA\ (panels (a1)-(a5)) and SUTRI 465~\AA\ (panels (b1)-(b5)). Panel (c) is superimposed by contours of the magnetic field at $\pm$ 400 Gauss levels, with yellow indicating positive values and blue representing negative values.
  \label{fig:Fig.7}}
\end{figure}

\begin{figure}
  \centering
  \includegraphics[width=1.0\textwidth]{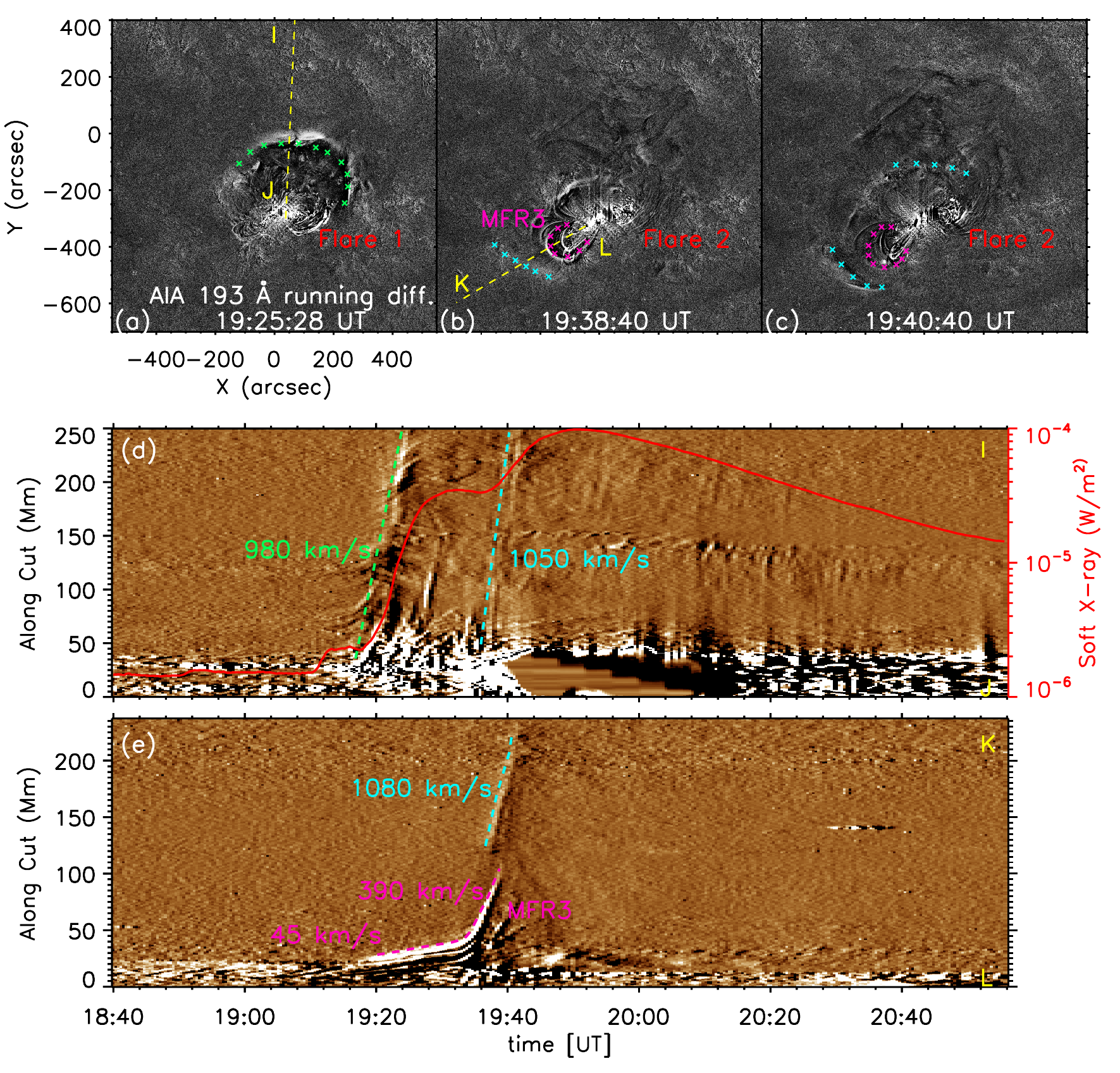}
  \caption{EUV waves produced by the eruption. Panels (a)-(c) shows the EUV wavefronts propagating outward the AR 13500 in AIA 193~\AA\ running difference with a 48-second cadence. The green and cyan cross signs indicate the observed wavefront, with the green one associated with Flare 1 (M3.4-class flare) and the others associated with Flare 2 (M9.8-class flare). The pink cross signs indicate the loops of MFR3. Yellow dashed lines are placed to generate of time-distance diagrams, as illustrated in panels (d)-(e). These slices are chosen to illustrate the kinetics of the EUV waves and MFR loops. Panels (d)-(e) shows the time-distance diagrams of AIA 193~\AA\ running difference. The green, cyan and pink dashed lines are corresponding to the fronts depicted in the same color in panels (a)-(c). Panel (d) is overlaid with the GOES SXR 1-8~\AA\ curve in red.
  \label{fig:Fig.8}}
\end{figure}

\begin{figure}
  \centering
  \includegraphics[width=1.0\textwidth]{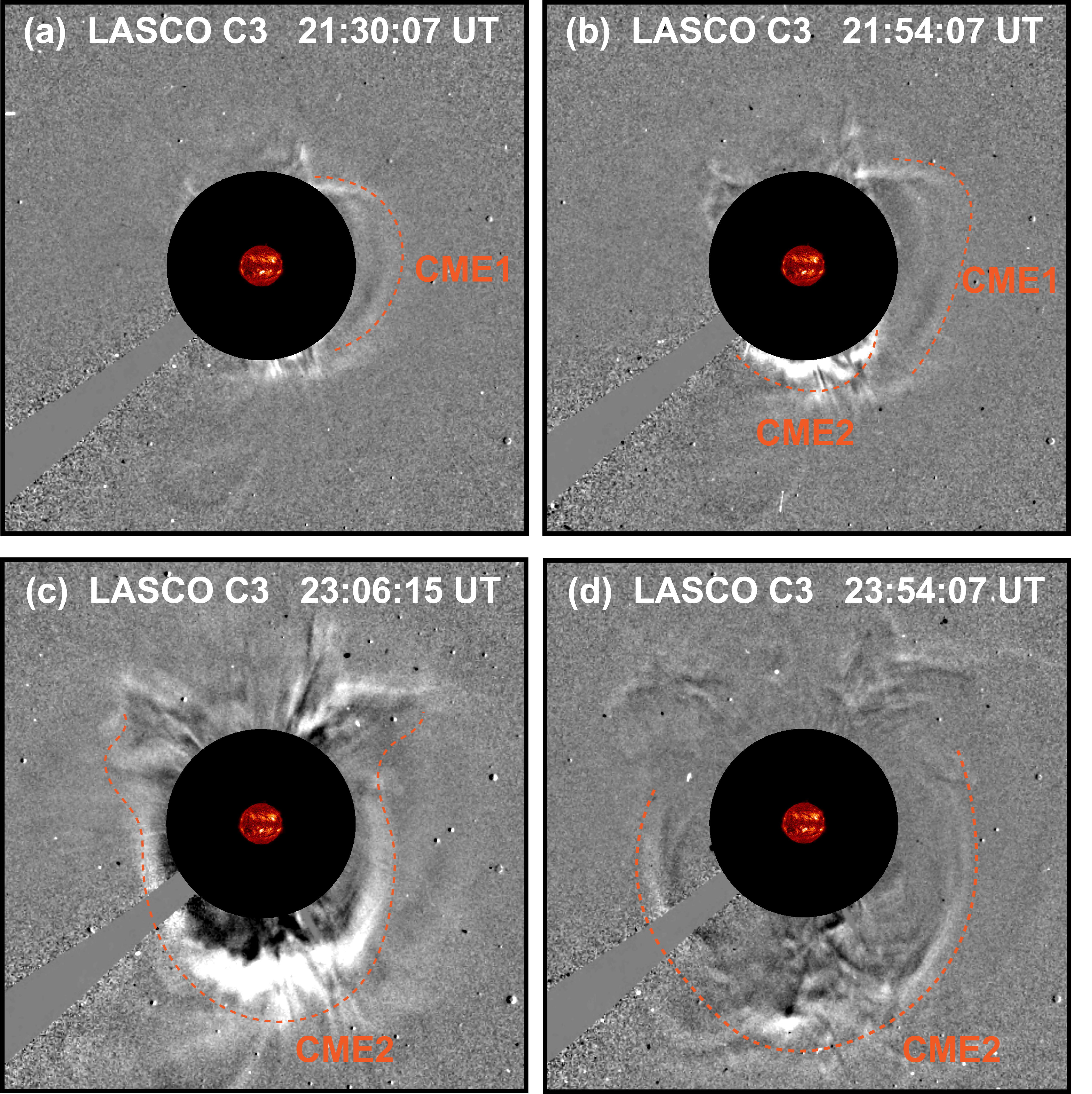}
  \caption{LASCO white-light observations within the near-Sun corona.
  \label{fig:Fig.9}}
\end{figure}

\begin{figure}
  \centering
  \includegraphics[width=1.0\textwidth]{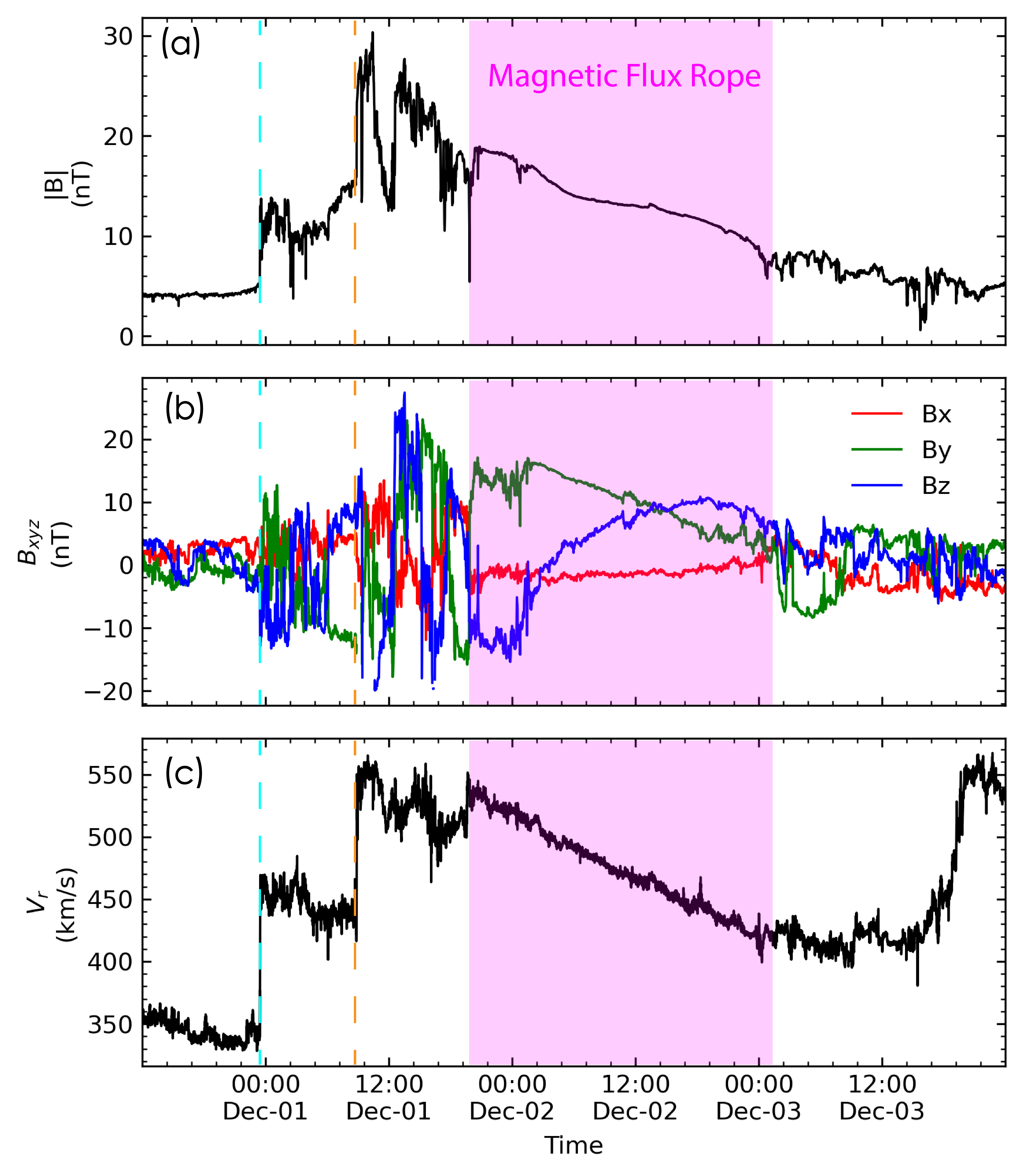}
  \caption{Wind in-situ observations of the ICME. The panels from top to bottom, display plots of magnetic field strength ($\left|B\right|$) and its components (Bx, By, Bz), as well as solar wind speed (Vr) as a function of time, respectively. The cyan and orange dashed lines indicate two instances of fast shock arrival, with the cyan line corresponding to the ICME produced one day before the eruption under analysis, and the orange line corresponding to the ICME being analyzed. The time period highlighted in pink denotes the duration of MFR passage.
  \label{fig:Fig.10}}
\end{figure}

\begin{figure}
  \centering
  \includegraphics[width=1.0\textwidth]{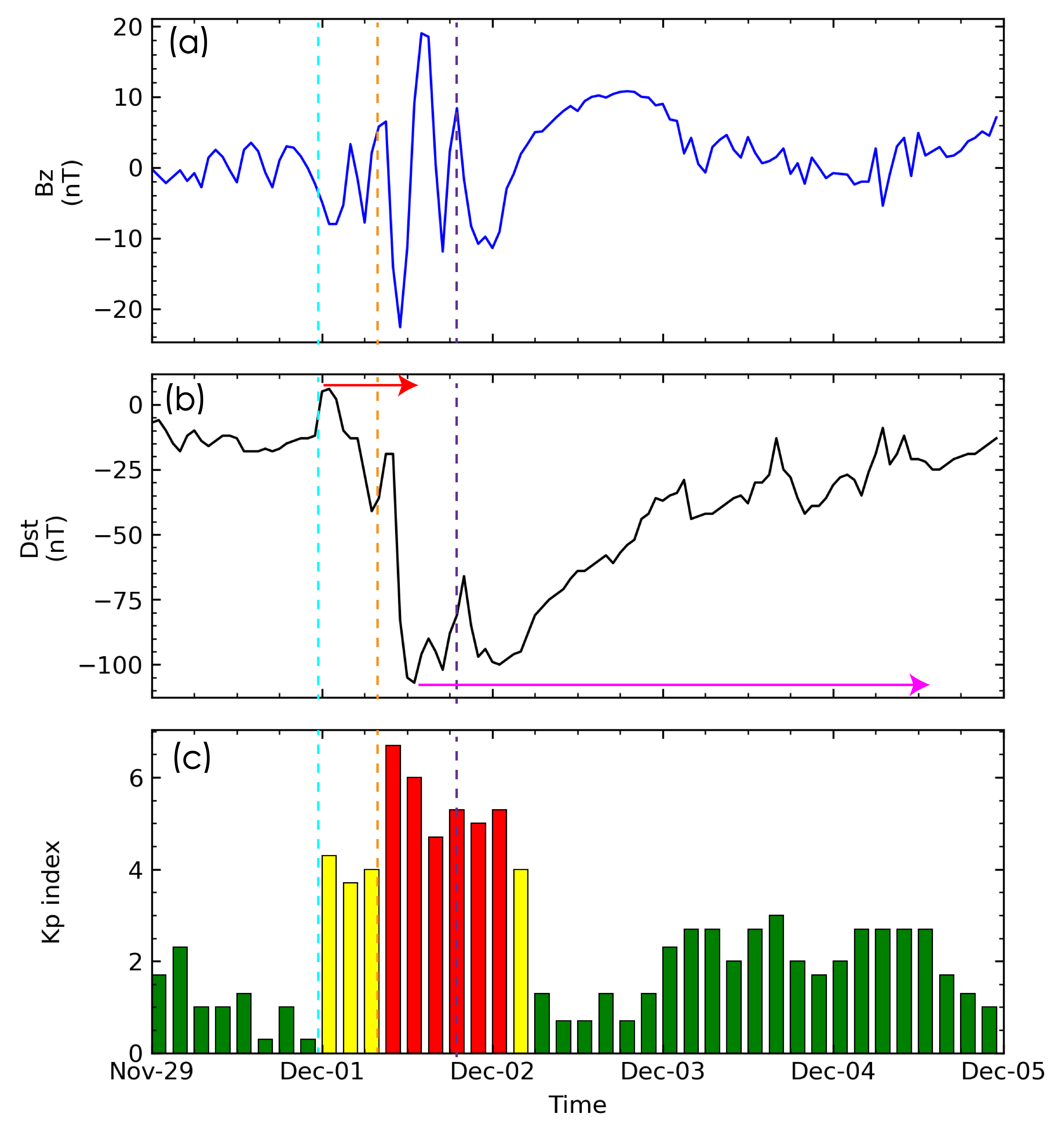}
  \caption{In-situ observations of the CME encountering Earth's atmosphere. Panel (a) presents the Bz in GSE coordinates averaged over one hour. Panel (b) displays the Dst index, reaching a minimum of -105 nT, indicating an intense geomagnetic storm. The cyan and orange dashed lines indicate the shock arrival times as in Fig. 10. The purple dashed line indicates the MFR arrival time, also shown in Fig. 10. The red and pink arrows denote the main phase and recovery phase of the geomagnetic storm, respectively. Panel (c) shows the histogram of the Kp index, with green, yellow, and red bars representing index values 0-3, 3-4.3, and 4.3-7, respectively. The Kp index values reach up to 6.7 during the storm's main phase, categorized as G2 according to NOAA Space Weather Scales.
  \label{fig:Fig.11}}
\end{figure}

\end{document}